\documentclass[tightenlines,twocolumn,amsmath,amssymb,showkeys,twoside,aps,floatfix,prb]{revtex4}

\usepackage{ifthen}
\newboolean{JoT}
\setboolean{JoT}{false}

\usepackage{amsmath} 
\usepackage{graphicx} 
\usepackage{dcolumn} 
\usepackage{bm} 
\usepackage{color} 
\usepackage{mathrsfs}
\usepackage{fancyhdr}
\usepackage[colorlinks,letterpaper,dvips]{hyperref}

\hypersetup{citecolor=blue,
            linkcolor=blue,
	    urlcolor=blue,
	    pdftitle=Exploring the beta-pdf in variable-density turbulent mixing,
	    pdfauthor=J. Bakosi and J. R. Ristorcelli,
	    pdfkeywords=Probability density function method; Variable-density turbulence; Transition to turbulence; Active scalar mixing; Beta distribution}
\usepackage{hypernat}

\begin{document}

\newcommand{\laur}{LA-UR 10-01595, v1.2, \emph{Accepted in Journal of Turbulence, July 19, 2010}}
\newcommand{\version}{\laur}

\newcommand{\bv}[1]{{\mbox{\boldmath$#1$}}} 
\newcommand{\fmean}[1]{{\left\langle{#1}\right\rangle}} 
\newcommand{\ifmean}[1]{{\langle{#1}\rangle}} 
\newcommand{\rmean}[1]{{\overline{#1}}} 
\newcommand{\rf}{'} 
\newcommand{\ff}{''} 
\newcommand{\rv}[1]{\rmean{#1\rf^2}} 
\newcommand{\fv}[1]{\fmean{#1\ff^2}} 
\newcommand{\rs}[1]{\rmean{#1\rf^3}} 
\newcommand{\rk}[1]{\rmean{#1\rf^4}} 
\newcommand{\ld}[1]{\frac{\mathrm{d}#1}{\mathrm{d}t}} 
\newcommand{\sd}[1]{\frac{\mathrm{D}#1}{\mathrm{D}t}} 
\newcommand{\msd}[1]{\frac{\mathrm{\overline{D}}#1}{\mathrm{\overline{D}}t}} 
\newcommand{\erf}[1]{\mathrm{erf}\left(#1\right)} 
\newcommand{\sm}{{\scriptstyle\mathcal{M}}} 

\newcommand{\vd}{d} 
\newcommand{\vdrf}{d\rf} 
\newcommand{\ivdrf}{d\rf} 

\newcommand{\Eqr}[1]{(\ref{#1})}
\newcommand{\Eqre}[1]{Eq.~(\ref{#1})}
\newcommand{\Eqrs}[1]{(\ref{#1})}
\newcommand{\Eqres}[1]{Eqs.~(\ref{#1})}
\newcommand{\Fig}[1]{figure~\ref{#1}}
\newcommand{\Figs}[1]{figures~\ref{#1}}
\newcommand{\Fige}[1]{Figure~\ref{#1}}
\newcommand{\Figse}[1]{Figures~\ref{#1}}
\newcommand{\fig}[1]{fig.~\ref{#1}}
\newcommand{\figs}[1]{figs.~\ref{#1}}
\newcommand{\fige}[1]{Fig.~\ref{#1}}
\newcommand{\figse}[1]{Figs.~\ref{#1}}

\ifthenelse{\boolean{JoT}}
{}
{\pagestyle{fancy}
 \fancyhead{}
 \fancyhead[LE,RO]{\thepage}
 \fancyfoot{}
 \chead{\texttt{\version}}
 \renewcommand{\headrulewidth}{0pt}}

\ifthenelse{\boolean{JoT}}
{\title{Exploring the beta-pdf in variable-density turbulent mixing}}
{\title{Exploring the beta-pdf in variable-density turbulent mixing\\\small\texttt{\version}}}

\ifthenelse{\boolean{JoT}}
{\author{J. Bakosi and J. R. Ristorcelli$^{\ast}$\thanks{$^\ast$Email: \{jbakosi,jrrj\}@lanl.gov\vspace{6pt}}\\
\vspace{6pt}{\em{Los Alamos National Laboratory, Los Alamos, NM 87545, USA}}\\
\vspace{6pt}\received{\version}}}
{\author{J. Bakosi}\author{J. R. Ristorcelli\\\small\texttt{\{jbakosi,jrrj\}@lanl.gov}}\affiliation{Los Alamos National Laboratory, Los Alamos, NM 87545, USA}}

\ifthenelse{\boolean{JoT}}
{\maketitle}
{}

\begin{abstract}
In assumed probability density function (pdf) methods of turbulent combustion, the shape of the scalar pdf is assumed \emph{a priori} and the pdf is parametrized by its moments for which model equations are solved. In non-premixed flows the beta distribution has been a convenient choice to represent the mixture fraction in binary mixtures or a progress variable in combustion. Here the beta-pdf approach is extended to variable-density mixing: mixing between materials that have very large density differences and thus the scalar fields are active. As a consequence, new mixing phenomena arise due to 1) cubic non-linearities in the Navier-Stokes equation, 2) additional non-linearities in the molecular diffusion terms and 3) the appearance of the specific volume as a dynamical variable.

The assumed beta-pdf approach is extended to transported pdf methods by giving the associated stochastic differential equation (SDE). This enables the direct computation of the scalar pdf in a Monte-Carlo fashion. Using the moment equations, derived from the governing SDE, we derive constraints on the model coefficients of the SDE that provide consistency conditions for binary material mixing. The beta distribution is shown to be a realizable, consistent and sufficiently general representation of the marginal pdf of the fluid density, an active scalar, in non-premixed variable-density turbulent mixing. The moment equations derived from mass conservation are compared to the moment equations derived from the governing SDE. This yields a series of relations between the non-stationary coefficients of the SDE and the mixing physics. All rigorous mathematical consequences of assuming a beta-pdf for the fluid mass density.

Our treatment of this problem is general: the mixing is mathematically represented by the divergence of the velocity field which can only be specified once the problem is defined. A simple example of the wide range of physical problems is isobaric, isothermal, large-density binary material mixing. A more complex one is mixing and combustion of non-premixed reactants in which the divergence is related to the source terms in the energy and species conservation equations. In this paper we seek to describe a theoretical framework to subsequent applications. We report and document several rigorous mathematical results, necessary for forthcoming work that deals with the applications of the current results to model specification, computation and validation of binary mixing of inert fluids.
\end{abstract}

\ifthenelse{\boolean{JoT}}
{}
{}

\ifthenelse{\boolean{JoT}}
{}
{\maketitle}

\section{Introduction}

In turbulent flows density fluctuations may arise due to non-uniform species concentrations, temperature or pressure. We concentrate here on the first case, where differences in the fluid mass density (e.g.\ due to mixing of different-density species) are very large and play a crucial role in the developing flow. To distinguish from classical constant-density shear-driven turbulence, we call the resulting turbulent flow \emph{variable-density} (VD) \emph{pressure-gradient-driven turbulence} (PGDT) due to the active and important role played by the density fluctuations in the coupled hydrodynamics and mixing processes.

A canonical example of PGDT is the Rayleigh-Taylor (RT) instability of an interface between two fluids of different densities, which occurs when an external acceleration force is directed opposite to the density gradient \cite{Rayleigh_82,Taylor_50,Sharp_84}. This phenomenon is present in terrestrial examples (e.g.\ in the unstable atmospheric boundary layer), in astrophysics (in a collapsed core of a massive star), as well as in engineering (in laser-driven or electromagnetic fusion). PGDT phenomena also occur in mixtures of different-density species accelerated by large pressure gradients, as in supersonic injectors, gas turbines or scramjet combustion in hypersonic vehicles.

Many engineering combustion simulations employ either \emph{assumed} or \emph{transported} probability density function (pdf) methods to compute the scalar mixing fields \cite{Fox_03}. In assumed pdf methods, the shape of the pdf of certain material fields is assumed \emph{a priori} and their distribution is parametrized by solving for appropriate moments. In contrast, transported pdf methods integrate the modelled evolution equations of the pdf \cite{Pope_85}. The main advantage of these methods, compared to moment closures, is the mathematically exact and closed form of the chemical source term, which eliminates the need for closure assumptions for such highly non-linear processes. Another advantage of pdf methods is that they provide a higher level statistical description of the turbulent fields.

Although the beta distribution had been widely employed before, Girimaji \cite{Girimaji_91} appears to have been the first to rigorously discuss it as a model for the pdf of turbulent mixing of inert passive scalars. Due to the lack of experimental data and the only direct numerical simulation (DNS) data \cite{Eswaran_88} on two-scalar mixing at the time, his proposal was limited to mixing of passive scalars in stationary homogeneous isotropic constant-density turbulence.

The recent DNS data of Livescu \& Ristorcelli \cite{Livescu_08} suggests that the beta-pdf may also be an appropriate model for active scalar mixing in \emph{variable-density, buoyancy-} or equivalently, \emph{pressure-gradient-driven} turbulence \cite[see also][and references therein]{Livescu_09}. The analysis of Livescu and Ristorcelli treat the mixing of different-density fluids with an external acceleration force (e.g.\ gravity). This is also the driver of the Rayleigh-Taylor instability. In RT flows, small initial perturbations of the interface between the fluids rapidly become a fully turbulent flow. The simulations analysed by Livescu and Ristorcelli are of a homogeneous RT flow in which the mean density field is constant; the simulations were designed to study the issues of 1) transition, 2) non-equilibrium and 3) what new phenomena might arise when variable-density effects are important.

\textbf{Mixing in Boussinesq turbulence.} If the densities of two fluids undergoing mixing are commensurate, the momentum equation for a buoyancy-driven flow has the form of a Boussinesq fluid: the density fluctuation is small compared to the mean density and important only in the body force, and the pressure gradient is the static head. The second moment equations for the Boussinesq fluid are given by Ristorcelli \& Clark \cite{Ristorcelli_Clark_04}. In the mixing of two pure fluids, the pdf of the fluid density proceeds from a double-delta configuration to a single-peaked Gaussian-like distribution at late time. Given symmetric initial conditions the pdf is symmetric at all later times; the skewness is zero at all times in the mixing of a (homogeneous) Boussinesq fluid if the skewness is zero at the outset.

\textbf{Mixing in variable-density turbulence.} If the density of the fluids are vastly different, the phrase \emph{variable-density (VD)} is used to distinguish from the Boussinesq case. In VD flows the pressure field applied to the very-different-density fluids gives rise to differential fluid accelerations and a variety of inertial effects (added mass effects being one of them). The advection term in the Navier-Stokes equation then contains cubic non-linearities, resulting in several non-Boussinesq features. For example, the mixing process becomes highly asymmetric as an initially symmetric pdf rapidly develops a sizeable skewness depending on the density differences of the two pure fluids. These issues, first found by Livescu \& Ristorcelli \cite{Livescu_08} are further discussed in Ref.\ \cite{Livescu_09b}. One of the consequences of these new effects is the dynamic importance of the specific volume and its correlations, required to close the hydrodynamical equations \cite{Livescu_09}. Similar correlations appear in the equations describing the mixing field and its moments. One of the purposes of this article, and the following papers on model development and application \cite{Bakosi_10b,Bakosi_10c}, is assessing the utility of an assumed beta-pdf in representing these new VD effects that appear as correlations with the fluctuating specific volume.

\textbf{Mixing in non-equilibrium turbulence.} A distinctive feature of the buoyancy-driven simulations of Livescu \& Ristorcelli \cite{Livescu_07,Livescu_08} is their transient nature. The flow evolution starts from a quiescent state, transitions to fully developed turbulence which is then followed by a final period of decay. The flow is highly non-equilibrium: the production-to-dissipation ratio, $\mathcal{P}/\epsilon$, before the transition at the inception of the flow is on the order of a few hundred; at the time of fully developed turbulence $\mathcal{P}/\epsilon \sim 1$ and in the final decay $\mathcal{P}/\epsilon \sim 0$. Consequently, a mixing model that is based on quasi-equilibrium assumptions is almost certainly in error at all stages of the flow evolution. Another purpose of this article is assessing the utility of the beta-pdf approach to representing not only the variable-density effects but also such non-stationary, non-equilibrium effects.

\subsection{Objectives of this article}
In this paper we explore and seek to resolve several issues of both modeling and theoretical nature. We derive the rigorous mathematical consequences of one simple (but reasonable) assumption in variable-density binary mixing: the statistical distribution of the fluid mass density is beta. Using this theoretical framework, and a coupled development on the stochastic momentum equation \cite{Bakosi_10b}, a subsequent paper \cite{Bakosi_10c} deals with the applications of the current results in model computations and validations. Here we seek:
\begin{enumerate}
\item To present a stochastic differential equation (SDE) for the beta distribution.
\item To establish the possibility and the requirements of a beta-pdf for modeling variable-density effects.
\item To establish the possibility and the requirements of a beta-pdf for modeling turbulence that has a highly transitional and non-equilibrium nature. This issue is further addressed in the sequels \cite{Bakosi_10b,Bakosi_10c}.
\item Two sets of moment equations are derived: one from the SDE yielding a beta distribution and one from exact mass conservation. These two sets of equations are then compared to determine what physical processes the parameters in the SDE are related to.
\item The two sets of moment equations are then compared to investigate the following question: If the density pdf were a beta distribution how would one consistently close very different terms in the moment equations required to describe VD mixing?
\item To explore what additional implications of an assumed mass density pdf might yield in polytropic media.
\end{enumerate}

Our primary focus will be to establish and document rigorous mathematical results regarding the above issues.

\subsection{Outline of the paper}
It seems useful to give an outline so that the mathematics does not obscure the purpose, direction and motivation of the presentation.

\begin{enumerate}
\item Sec.\ \ref{sec:context} lays the groundwork and background for the model development. Because of our interest in high-Schmidt-number mixing, the double-delta ``no-mix''-limit pdf is given. This limit is also of interest in early-time mixing of two pure fluids.
\item Sec.\ \ref{sec:beta} presents a stochastic evolution equation and shows that it yields a beta distribution at all times. Apparently, this SDE has not been given in the literature.
\item From the SDE the equations for the first several moments are derived in Sec.\ \ref{sec:moments}.
\item In Sec.\ \ref{sec:mass_consistency} we then establish that the SDE has the properties that allow it to represent the fluid mass density and to describe material mixing. The moment equations from the SDE are compared to the moment equations derived from mass conservation. This establishes relations between the parameters in the SDE and the physical mixing processes.
\item Up to this point the equations have addressed the statistically homogeneous flow, in which all the primary mixing mechanisms are seen. In Sec.\ \ref{sec:inhom} the equations for inhomogeneous flows are given and discussed.
\item Finally, in Sec.\ \ref{sec:conclusion} conclusions are drawn and the utility of the development is discussed.
\item The essential results, the model equation and its consistency conditions for material mixing, are summarized in Appendix \ref{app:sum}. Appendix \ref{app:pol} derives the consequences of a polytropic equation of state. The symmetric beta-pdf, a special case, is investigated in Appendix \ref{app:sym}.
\end{enumerate}

\section{Preliminary considerations}
\label{sec:context}
The context of the current work is given as a prelude to setting up the problem. Some basic results and issues relating to fluid mixing physics are summarized.

\textbf{The fluid mass density as a mixing variable.} Turbulent mixing of conserved scalars is a well-explored area in combustion theory and atmospheric pollution modeling. Both moment closures and pdf methods have been used to predict statistics (or the full pdf) of mass concentrations and mixture fractions. Mixing models have been developed for these quantities, because they are of the main interest in their applications. In flows with large density variations the mass density, $\varrho$, is an active scalar \cite{Livescu_07}, thus it is reasonable to develop a mixing model directly for this quantity. This is especially important for pressure-gradient-driven flows, as the product of the mass flux and the mean pressure gradient is an important source of turbulence and reflects the fact that Lagrangian particles of different-density fluids accelerate very differently in response to pressure gradients.

\textbf{The beta-pdf in variable-density mixing of a binary mixture.} In binary mixing between fluids of large density differences there is a one-to-one correspondence between the mass concentration and the fluid density. Due to the fact that the scalar is active and its fluctuations are not small, the momentum equation contains cubic non-linearities from the product of density and the quadratic velocity. For this reason it is more convenient in variable-density turbulence, from the viewpoint of the moment equations, to employ the fluid density as the mixing variable.

The beta distribution is a natural choice to represent the density in binary fluid flows since $\varrho$ can be expressed as a function of two \emph{dependent} passive scalars, as
\begin{equation}
\frac{1}{\varrho} = \frac{Y_1}{\varrho_1} + \frac{Y_2}{\varrho_2},
\end{equation}
where $\varrho_1$ and $\varrho_2$ denote the constant densities of the pure fluids and $Y_1 + Y_2 = 1$. This is the same situation as with the mixture fraction \cite{Bilger_89} in conditional moment closure (CMC) \cite{Klimenko_Bilger_99} methods and the two-scalar mixing of Girimaji \cite{Girimaji_91} in constant-density flows. In CMC the beta distribution is a convenient choice for representing the (assumed) pdf of the mixture fraction in chemically reacting turbulence. In that framework the reasons for the choice of the beta distribution are:
\begin{enumerate}
\item The mixture fraction is a conserved, bounded scalar representing the state of a two-component mixture ($Y_1$ and $1-Y_1$).
\item The ability to represent singular delta-function peaks of the unmixed state is required.
\item A continuous (even at the singularities) and always integrable pdf is required.
\item In the CMC equations the full pdf of the mixture fraction is required.
\item The physical pdf in the mixing of a binary non-premixed flow attains a wide variety of shapes, which can be well-approximated by the beta distribution.
\end{enumerate}

Similarly, in many classes of combustion problems in which the thermodynamic pressure can be assumed large and constant, and in which a progress variable, $0<c<1$, is used to describe the mixing and combustion process, one has
\begin{equation}
\varrho = \frac{\varrho_1 T_1}{T} = \frac{1}{1+\tau c},
\end{equation}
with the mixture temperature, $T$, the temperature of species ``1'', $T_1$, and the heat-release parameter, $\tau$. Then one can use the beta-pdf for the density field \cite{Bilger_80}.

Mixing in these classes of flows is done by the fluctuating velocity divergence. In the binary mixing problem, for example, one has for the dilatation
\begin{equation}
v\rf_{i,i} = \vdrf = -D(\ln\varrho\rf),_{ii},
\end{equation}
see Ref.\ \cite{Livescu_07}. For the Boussinesq fluid problem, in which the fluctuating density is small compared to some constant mean density, $\varrho = \varrho_0  + \varrho\rf = \varrho_0 + \beta c$, the dilatation is $\ivdrf = -Dc\rf\!,_{jj}$. The dilatation in all these approximations can be derived from the equation of state, $\varrho = \varrho(P,T,Y)$ and continuity. It is more complicated for the combustion problem, depending on various additional approximations that are made. For the constant thermodynamic pressure case one has
\begin{equation}
\vdrf = -\ld{}\ln\varrho(T,Y) + \rmean{\ld{}\ln\varrho(T,Y)}.
\end{equation}
In this case the dilatation, which accomplishes the molecular mixing, is most generally carried as the primary variable. In our treatment, in order to treat the general variable-density problem, the fluctuating dilatation is carried as the primary mixing variable.

\subsection{The density evolution in buoyant mixing}
\label{sec:features}
The time evolution of the density pdf during homogeneous buoyantly-driven variable-density mixing is shown in Fig.\ 4 of Ref.\ \cite{Livescu_08}, and reveals several unique features:
\begin{enumerate}

\item \emph{The initial double-delta pdf of the unmixed state:} In DNS of homogeneous RT mixing the computational domain consists of random blobs of two fluids with different densities at rest initially, the pdf is double-delta. This is the same distribution also in the inhomogeneous RT layer at the centreline at early time, representing unmixed quiescent fluids \cite{Livescu_09}.

\item \emph{Molecular mixing with persistent extrema:} With time the initial delta-peaks diffuse into a series of distributions whose variance is decreased by the increasing occurrence of density events in the  interior of the sample space. During this time there is still substantial unmixed fluid and the pdf is still bimodal but with two different peaks at the stationary extrema of the pure fluids. Only a small fraction of the ensemble corresponds to mixed fluid. This behaviour of the pdf \emph{cannot be represented} by simple deterministic relaxation models, widely used for passive scalar mixing, such as the \emph{interaction by exchange with the mean} (IEM) model \cite{Villermaux_Devillon_72,Dopazo_OBrien_74} and its variants. These models relax the \emph{full} ensemble, i.e.\ all scalar fluctuations, towards the mean on the \emph{same} time-scale, without accounting for changes in the scalar pdf, simply by reducing the distance between the delta peaks and leaving the shape intact. Relaxing only a fraction of the ensemble on possibly different time-scales (thereby influencing the shape of the evolving distribution) requires a stochastic model.

\item \emph{Mixing asymmetry:} Peculiar to VD mixing is the fact that the two different delta-function peaks do not decrease in magnitude at the same rate: the heavy fluid mixes molecularly more slowly than the light fluid. This is due to inertial effects and the ability of the heavy fluid to resist deformation, which induces a skewness into the distribution, most noticeable at higher initial density ratios. This is, in part, responsible for the bubble/spike asymmetry in turbulent Rayleigh-Taylor layers \cite{Livescu_09b}.

\item \emph{The distribution becomes asymptotically Gaussian at late times:} At later times the initially convex-shaped distribution evolves into concave shapes, and eventually the pdf arrives at an approximate  (clipped) Gaussian close to a fully mixed state with an ever-decreasing variance.
\end{enumerate}

The behaviour of the physical pdf described above can not be reproduced with simple Gaussian mixing models or by deterministic relaxation models. To capture this behaviour, it appears, \emph{a fundamentally different representation is required.}

\subsection{The no-mix double-delta distribution}
\label{sec:delta}
It is useful to have some asymptotic results. The various moment relations for the double-delta distribution are now derived and summarized.

The double-delta distribution, with delta-peaks at $\varrho_1>0$ and $\varrho_2>0$, with the asymmetry-parameter $0 \le S \le 1$, is defined as
\begin{equation}
\mathscr{D}(\varrho) = (1-S)\Hat{\delta}(\varrho-\varrho_1) + S\Hat{\delta}(\varrho-\varrho_2),
\end{equation}
where $\Hat{\delta}(\varrho)$ denotes the Dirac-delta function. Then the mean, $\rmean{\varrho}$, is given by
\begin{equation}
\rmean{\varrho} \equiv \int \varrho \mathscr{D}(\varrho)\mathrm{d}\varrho = (1-S)\varrho_1 + S\varrho_2,
\end{equation}
while the central moments for $n\ge2$ are
\begin{equation}
\rmean{\varrho\rf^n} \equiv \int (\varrho-\rmean{\varrho})^n\mathscr{D}(\varrho)\mathrm{d}\varrho = (1-S)(\varrho_1-\rmean{\varrho})^n + S(\varrho_2-\rmean{\varrho})^n.
\end{equation}
The mean of $v=1/\varrho$ is
\begin{equation}
\rmean{v} \equiv \int \frac{1}{\varrho} \mathscr{D}(\varrho)\mathrm{d}\varrho = (1-S)v_1 + Sv_2,
\end{equation}
with $v_1=1/\varrho_1$ and $v_2=1/\varrho_2$ and the central moments of $v$, for $n\ge2$, are
\begin{equation}
\rmean{v\rf^n} \equiv \int \left(\frac{1}{\varrho}-\rmean{v}\right)^n \mathscr{D}(\varrho)\mathrm{d}\varrho = (1-S)(v_1-\rmean{v})^n + S(v_2-\rmean{v})^n.
\end{equation}
Finally, the covariance of $\varrho$ and $v$ becomes
\begin{align}
\rmean{\varrho\rf v\rf} & \equiv \int (\varrho-\rmean{\varrho})\left(\frac{1}{\varrho}-\rmean{v}\right) \mathscr{D}(\varrho)\mathrm{d}\varrho = 1 - \rmean{\varrho}\!\cdot\!\rmean{v} = \nonumber\\
& = 1- \left[S^2 + \left(\frac{\varrho_1}{\varrho_2}+\frac{\varrho_2}{\varrho_1}\right)S(1-S) + (1-S)^2\right],
\end{align}
which shows that $\rmean{\varrho\rf v\rf}<0$ since $(\varrho_1/\varrho_2+\varrho_2/\varrho_1) \ge 2$.

Based on the above, the mean, variance, skewness and kurtosis of $\varrho$, in the special case of $\varrho_1=0$ and $\varrho_2=1$, are given by
\begin{align}
\rmean{\varrho} & = S,\label{eq:dd1}\\
\rv{\varrho} & = S(1-S),\label{eq:dd2}\\
\frac{\rs{\varrho}}{\rv{\varrho}^{3/2}} & = \frac{1-2S}{\sqrt{S(1-S)}},\label{eq:dd3}\\
\frac{\rk{\varrho}}{\rv{\varrho}^2} & = \frac{1}{S(1-S)} - 3.\label{eq:dd4}
\end{align}

\section{The beta-pdf model}
\label{sec:beta}
This section presents a SDE that yields a beta distribution. Our intention is to develop a model to represent the fluid density field and its moments in transitional, variable-density, pressure-gradient-driven turbulent mixing with molecular diffusion. The formulation will prove useful in more specialized cases, such as passive two-scalar mixing with equal (or unequal) diffusivity, or asymmetric mixing with negligible molecular diffusion.

\subsection{The beta distribution}
\begin{figure}
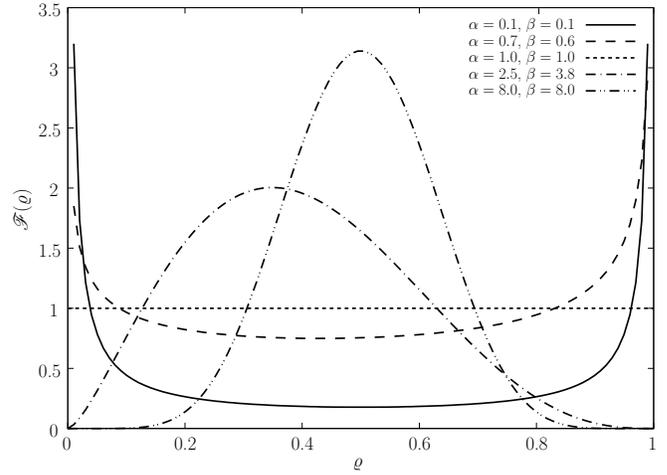

\centering
\ifthenelse{\boolean{JoT}}
{\resizebox{0.7\columnwidth}{!}{\input{beta-distributions.pstex_t}}}
{\resizebox{\columnwidth}{!}{\input{beta-distributions.pstex_t}}}
\caption{Beta distributions with different parameters $\alpha$ and $\beta$. If $\alpha\!=\!\beta$ the distribution is symmetric, while $\beta\!>\!\alpha$ gives positive and $\alpha\!>\!\beta$ gives negative skewness, respectively. For $\alpha\!>\!1$ and $\beta\!>\!1$ the distribution has a single peak, whereas $\alpha\!<\!1$ and $\beta\!<\!1$ corresponds to a U-shaped beta distribution.}
\label{fig:beta}
\end{figure}

The \emph{beta distribution} (\fige{fig:beta}) with parameters $\alpha>0$ and $\beta>0$ is
\begin{equation}
\mathscr{F}(\varrho) = \frac{\varrho^{\alpha-1}(1-\varrho)^{\beta-1}}{B(\alpha,\beta)},
\label{eq:beta}
\end{equation}
where $B(\alpha,\beta)$ denotes the value of the \emph{Euler beta function}
\begin{equation}
B(\alpha,\beta) = \int_0^1 y^{\alpha-1}(1-y)^{\beta-1}\mathrm{d}y,
\end{equation}
which ensures that the total probability $\mathscr{F}(\varrho)$ integrates to unity.

This distribution is bounded between $0 \le\varrho\le 1$ with a possible singular behaviour at the extremes depending on $\alpha$ and $\beta$, enabling the representation of the delta-function peaks in the initial unmixed state. The boundedness for all values of $\alpha$ and $\beta$ ensures that in a transported pdf method no fluid particles can have densities lower or higher than the initial densities, which is the correct behaviour if compressibility effects are negligible.

In \fige{fig:beta} the beta distribution is shown to possess the correct attributes for representing the density in VD-RT flows. The function can evolve from an initially convex to a concave shape through a series of possibly skewed distributions, incorporating mixing asymmetry due to VD effects, c.f.\ Fig.\ 4 in Ref.\ \cite{Livescu_08}.

The first two moments of the beta distribution, defined by \Eqre{eq:beta}, are
\begin{align}
\rmean{\varrho} & \equiv \int \varrho \mathscr{F}(\varrho)\mathrm{d}\varrho = \frac{\alpha}{\alpha+\beta},\label{eq:r-mean}\\
\rv{\varrho} & \equiv \int (\varrho-\rmean{\varrho})^2 \mathscr{F}(\varrho)\mathrm{d}\varrho = \frac{\alpha\beta}{(\alpha+\beta)^2(\alpha+\beta+1)}.\label{eq:r-variance}
\end{align}
As is well known, the beta distribution is fully determined by its mean and variance.

Based on \Eqres{eq:r-mean} and \Eqrs{eq:r-variance}, the two parameters, $\alpha$ and $\beta$, can be expressed in terms of the first two moments as
\begin{align}
\alpha & = \frac{\rmean{\varrho}}{\rv{\varrho}}\left[\rmean{\varrho}(1-\rmean{\varrho})-\rv{\varrho}\right] = \rmean{\varrho}\frac{\theta}{1-\theta},\\
\beta & = \frac{1-\rmean{\varrho}}{\rv{\varrho}}\left[\rmean{\varrho}(1-\rmean{\varrho})-\rv{\varrho}\right] = (1-\rmean{\varrho})\frac{\theta}{1-\theta},
\end{align}
which shows that the knowledge of $\rmean{\varrho}$ and $\rv{\varrho}$ determines all moments of the beta-pdf. Here
\begin{equation}
\theta = 1 - \frac{\rv{\varrho}}{\rmean{\varrho}(1-\rmean{\varrho})},
\end{equation}
is the commonly used mix-metric \cite{Livescu_08}. $\theta=0$ represents pure unmixed fluids, while $\theta=1$ occurs in the fully mixed state.

The above development re-iterated some of the well-known mathematical characteristics of the beta distribution and qualitatively justified its use as a suitable (and sufficiently general) representation of the fluid density pdf in variable-density RT flows.

\subsection{The stationary stochastic differential equation}
This section presents the stochastic equation that yields a beta-pdf at all times, together with its equivalent Fokker-Planck equation (FPE).

The stationary solution of the It\^o SDE governing the Lagrangian particle property, $0\le\varrho^*\le1$,
\begin{equation}
\mathrm{d}\varrho^*(t) = \frac{b}{2}(S-\varrho^*)\mathrm{d}t + \sqrt{\kappa\varrho^*(1-\varrho^*)}\mathrm{d}W(t),
\label{eq:dSDE}
\end{equation}
with the Wiener process $\mathrm{d}W(t)$, Ref.\ \cite{Gardiner_04}, and constant parameters $b>0$, $\kappa>0$ and $0 < S < 1$ is the beta distribution $\mathscr{F}(\varrho)$. This can be readily seen if \Eqre{eq:beta} is substituted into the FPE, equivalent to (and derived from) the SDE \Eqrs{eq:dSDE}, see e.g.\ Ref.\ \cite{Gardiner_04},
\begin{equation}
\frac{\partial \mathscr{F}}{\partial t} = -\frac{\partial}{\partial\varrho}\left[\frac{b}{2}(S-\varrho)\mathscr{F}\right] + \frac{1}{2}\frac{\partial^2}{\partial\varrho^2}\big[\kappa\varrho(1-\varrho)\mathscr{F}\big]
\label{eq:dFP}
\end{equation}
in the stationary limit, $\partial\mathscr{F}/\partial t = 0$, with all constant parameters
\begin{equation}
\alpha=S\frac{b}{\kappa} \quad \mathrm{and} \quad \beta=(1-S)\frac{b}{\kappa}.\label{eq:ab}
\end{equation}
\Eqre{eq:ab} shows that the SDE \Eqrs{eq:dSDE} with the three coefficients, $b$, $S$ and $\kappa$, in fact represents a two-parameter distribution.

The same result can be obtained by evaluating the integral in the stationary solution for \Eqre{eq:dFP}
\begin{equation}
\mathscr{F}(\varrho) = \frac{\mathcal{N}}{\kappa\varrho(1-\varrho)}\exp{\left(\frac{b}{\kappa}\int_{u=0}^\varrho\frac{S-u}{u(1-u)}\mathrm{d}u\right)},
\end{equation}
where $\mathcal{N}$ is a normalization constant suitable chosen so that $\mathscr{F}$ integrates to unity.

Note that the SDE \Eqrs{eq:dSDE} is a special case of
\ifthenelse{\boolean{JoT}}
{
\begin{equation}
\mathrm{d}\Hat{\varrho}^*(t) = \frac{b}{2}\left[S(\varrho_2-\varrho_1)+\varrho_1-\Hat{\varrho}^*\right]\mathrm{d}t + \sqrt{\kappa(\Hat{\varrho}^*-\varrho_1)(\varrho_2-\Hat{\varrho}^*)}\mathrm{d}W(t),\label{eq:dSDEr1r2}
\end{equation}
}
{
\begin{align}
\begin{split}
\mathrm{d}\Hat{\varrho}^*(t) & = \frac{b}{2}\left[S(\varrho_2-\varrho_1)+\varrho_1-\Hat{\varrho}^*\right]\mathrm{d}t\\
&\quad + \sqrt{\kappa(\Hat{\varrho}^*-\varrho_1)(\varrho_2-\Hat{\varrho}^*)}\mathrm{d}W(t),\label{eq:dSDEr1r2}
\end{split}
\end{align}
}
for $\varrho_1\le\Hat{\varrho}^*\le\varrho_2$. By setting $\varrho_1=0$ and $\varrho_2=1$, \Eqre{eq:dSDEr1r2} reduces to \Eqre{eq:dSDE}. For simplicity we will use \Eqre{eq:dSDE}, noting that the transformation
\begin{equation}
\Hat{\varrho}^* = \frac{1}{2}(\varrho_1+\varrho_2) + \left(\varrho^*-\frac{1}{2}\right)(\varrho_2-\varrho_1),
\end{equation}
yields the more general sample space bounds. It is worth keeping in mind that the ``$1$'' in the diffusion term of \Eqre{eq:dSDE} thus has the unit of $\varrho$. It can be seen from \Eqre{eq:dSDEr1r2} that $S$ is non-dimensional, while $b$ and $\kappa$ are inverse time-scales. \fige{fig:beta} with \Eqre{eq:ab} shows that $S$ is responsible for the skewness and the ratio
\begin{equation}
\delta=\frac{\kappa}{b}
\end{equation}
controls the convexity. For $S/\delta\!<\!1$ and $(1\!-\!S)/\delta\!<\!1$ the shape is convex, while $S/\delta\!>\!1$ and $(1\!-\!S)/\delta\!>\!1$ result in a concave shape.

\subsection{The non-stationary stochastic differential equation}
The stationary SDE \Eqrs{eq:dSDE} is now extended to non-stationary processes.

\Eqre{eq:dSDE} represents a time-homogeneous process, since its drift and diffusion coefficients are not explicit functions of time \cite{Gardiner_04}. In such a case, as shown above, the solution of the equation is a stationary distribution, $\mathscr{F}(\varrho)$, whose statistics are time-independent and the stochastic process $\varrho^*(t)$ is statistically stationary.

In the non-stationary case the coefficients, $b(t)$, $S(t)$ and $\kappa(t)$, are time-dependent and
\begin{equation}
\mathrm{d}\varrho^*(t) = \frac{b(t)}{2}(S(t)-\varrho^*)\mathrm{d}t + \sqrt{\kappa(t)\varrho^*(1-\varrho^*)}\mathrm{d}W(t).
\label{eq:dSDEt}
\end{equation}
\Eqre{eq:dSDEt} represents a beta distribution whose shape evolves in time. The time-dependent model coefficients govern the evolution of the shape of the pdf $\mathscr{F}[\varrho;b(t),S(t),\kappa(t)]$, abbreviated as $\mathscr{F}(\varrho;t)$. The FPE equivalent to \Eqre{eq:dSDEt} is formally the same as \Eqre{eq:dFP} but with time-dependent coefficients, $b(t)$, $S(t)$ and $\kappa(t)$, and its solution is $\mathscr{F}(\varrho;t)$.

The above development shows that \Eqre{eq:dSDEt} is a generalization of \Eqre{eq:dSDE} to a transient description. Consequently, special cases of stationary distributions can be obtained at any instant in time by setting all three coefficients to constant values, in which case \Eqre{eq:dSDEt} will reduce to \Eqre{eq:dSDE}. The solution, i.e.\ the pdf, will converge to a stationary distribution, $\mathscr{F}(\varrho;t)\to\mathscr{F}_\mathrm{s}(\varrho)$, whose shape is determined by the constants $b_\mathrm{s}$, $S_\mathrm{s}$ and $\kappa_\mathrm{s}$. We will use this property of \Eqre{eq:dSDEt} to investigate the effects of the model coefficients on the shape of the pdf and its statistics. It is worth emphasizing that $\mathscr{F}_\mathrm{s}(\varrho)$ is a stationary distribution in the mathematical sense, to which the solution of the SDE \Eqrs{eq:dSDEt} converges with constant coefficients. This state is not to be confused with the physical state of the modelled flow, which may be statistically non-stationary at any given time.

It is important to appreciate that $b(t)$, $S(t)$ and $\kappa(t)$ must be independent of $\varrho^*$, but they can be a function of any available single-point moments of the flow. For instance, $b[\rmean{\varrho}(t),\rv{\varrho}(t),\rmean{\varrho\rf v\rf}(t),\rmean{\varrho\rf\ivdrf}(t),\dots,t]$, which is abbreviated by $b(t)$. Thus in developing suitable specifications for $b$, $S$ and $\kappa$, all available one-point statistics of the flow can be employed.

We showed that a temporally evolving beta-pdf can be represented by a Lagrangian SDE \Eqrs{eq:dSDEt} with time-varying coefficients. This establishes the possibility of modeling statistically transitional, non-equilibrium flows and is a crucial ingredient of the joint pdf model developed in Refs.\ \cite{Bakosi_10b,Bakosi_10c} for the highly transitional and non-equilibrium RT turbulence.

\subsection{Stochastic equations in pdf methods}
The mathematical properties of the different types of SDEs in pdf methods are used to put \Eqre{eq:dSDEt} in a historical context.

From a mathematical viewpoint, the governing equations, employed by transported pdf methods, written as a general It\^o diffusion process for the scalar $y^*$,
\begin{equation}
\mathrm{d}y^*(t) = A(y^*,t)\mathrm{d}t + \sqrt{B(y^*,t)}\mathrm{d}W(t),
\end{equation}
can be categorized as:
\begin{enumerate}
\item \emph{Deterministic equation with $B(y^*,t) = 0$:} such as relaxation models for molecular mixing, including the IEM-family \cite{Villermaux_Devillon_72,Dopazo_OBrien_74},
\item \emph{Stochastic equation with constant diffusion, $B(y^*,t) = B(t)$:} such as the Langevin model for the velocity \cite{Haworth_86}, or the log-normal model for the kinetic energy dissipation time-scale \cite{Pope_90b},
\item \emph{Stochastic equation with linear diffusion, $B(y^*,t) = B(ay^*+b,t)$:} such as the gamma distribution model for the turbulence frequency \cite{vanSlooten_98}, and,
\item \emph{Stochastic equation with non-linear diffusion, $B(y^*,t)$:} such as the Fokker-Planck model for differential diffusion \cite{Fox_92,Fox_99}, or \Eqre{eq:dSDEt}.
\end{enumerate}
The drift is linear, $A(y^*,t) = A(ay^*+b,t)$, in all these equations.

Deterministic equations are incapable of influencing the shape of the evolving pdf, a highly desired property of mixing models \cite{Fox_03}. Equations with constant diffusion and linear drift yield a Gaussian pdf. This is suitable for an unbounded quantity (e.g.\ velocity, log-frequency), but unphysical for mixing. Equations with linear drift and diffusion give a gamma distribution, supported on a semi-infinite interval.

\Eqre{eq:dSDEt} belongs to the 4th group. The diffusion term in \Eqre{eq:dSDEt} is a function of the sample space, $\varrho^*$, and it effects a non-linear mapping of the Gaussian Wiener process. The non-linear diffusion coefficient, $\sqrt{\kappa\varrho^*(1-\varrho^*)}$, allows the SDE 1) to confine the process to a bounded interval on its sample space and 2) to influence the shape of its solution in a wide variety of ways. This allows \Eqre{eq:dSDEt} to capture some fundamentally different mixing phenomena that simple micro-mixing models, such as the IEM family, cannot reproduce.

Apparently, \Eqre{eq:dSDEt} has not appeared in the literature in this simple yet general form. Fox's \cite{Fox_92,Fox_03} treatment of differential diffusion is general enough to include the beta-pdf, though he works with the symmetric distribution, a special case of our treatment, where the drift relaxes to the centre of the sample space instead of $S(t)$. Cai \& Lin \cite{Cai_96} also give the symmetric equation in the stationary case. As is shown, in general \Eqre{eq:dSDEt} represents a non-stationary skewed beta-pdf.

The next section examines the three SDE coefficients, $b(t)$, $S(t)$ and $\kappa(t)$, that together determine $\alpha(t)$ and $\beta(t)$, in more detail.

\section{Moment equations from the SDE}
\label{sec:moments}
In order to identify possible constraints on the three model coefficients, $b(t)$, $S(t)$ and $\kappa(t)$, so that \Eqre{eq:dSDEt} can correctly represent material mixing, evolution equations for the first few statistical moments of $\mathscr{F}(\varrho;t)$ are now derived from \Eqre{eq:dFP}.

Balance equations for statistics are derived as follows. Multiplying each term in \Eqre{eq:dFP} with $\varrho$ and integrating over all sample space
\begin{equation}
\int_0^1 \varrho \frac{\partial\mathscr{F}}{\partial t} \mathrm{d}\varrho = \dots,
\end{equation}
produces the governing equation for the mean density, $\rmean{\varrho}(t)$. The equation for the density variance, $\rv{\varrho}(t)$, is obtained by multiplying with $(\varrho-\rmean{\varrho})^2$ and integrating each term. For mathematical details on evaluating the integrals the reader is referred to Ref.\ \cite{Pope_85}.

For clarity, the time-dependence of the model coefficients and the derived moments are not explicitly stated in the following but implied.

\subsection{Mean: $\rmean{\varrho}$}
Multiplying \Eqre{eq:dFP} by $\varrho$ and integrating over the interval $0\le\varrho\le1$ gives the governing equation for the time evolution of the mean, $\rmean{\varrho}$, of the stochastic process, $\varrho^*(t)$, governed by \Eqre{eq:dSDEt} as
\begin{equation}
\frac{\partial\rmean{\varrho}}{\partial t} = \frac{b}{2}(S-\rmean{\varrho}),\label{eq:dmean}
\end{equation}
showing that the stationary mean is
\begin{equation}
\rmean{\varrho}_\mathrm{s} = S,
\label{eq:sdmean}
\end{equation}
where the subscript $s$, as before, denotes the stationary value, $\partial/\partial t=0$. Thus \Eqre{eq:sdmean} shows that specifying $S$ in any way means specifying the stationary value, $\rmean{\varrho}_\mathrm{s}$, to which the mean of the distribution will converge since $b>0$.

Examining the statistics of the stationary state, such as that of $\rmean{\varrho}$ in \Eqre{eq:sdmean}, is helpful in determining the effect of the coefficients independent of time: it answers the question \emph{``what would be the shape of the pdf for given $b$, $S$ and $\kappa$?''} Since the stationary moment is an explicit function of only the coefficients (and only implicitly of time), it clearly indicates the effects of $b$, $S$ and $\kappa$ on the given moment. On the right hand side of \Eqre{eq:sdmean}, $S$ is understood as $S_\mathrm{s}$, but for clarity, only the moment under consideration (such as $\rmean{\varrho}_\mathrm{s}$) is marked by $s$.

\subsection{Variance: $\rv{\varrho}$}
Multiplying \Eqre{eq:dFP} by $(\varrho-\rmean{\varrho})^2$ then integrating produces the evolution equation for the variance of the beta-pdf, governed by \Eqre{eq:dSDEt},
\begin{equation}
\frac{1}{b}\frac{\partial\rv{\varrho}}{\partial t} = \delta\rmean{\varrho}(1-\rmean{\varrho}) - (1+\delta)\rv{\varrho} = -\rmean{\varrho}(1-\rmean{\varrho})\big[1-\theta(1+\delta)\big],
\label{eq:dv}
\end{equation}
indicating that $\rv{\varrho}$ at any time will converge to its stationary value
\begin{equation}
\rv{\varrho}_\mathrm{s}=\frac{\delta}{1+\delta}S(1-S).\label{eq:sdv}
\end{equation}

Alternatively, the stationary moments, $\rmean{\varrho}_\mathrm{s}$, $\rv{\varrho}_\mathrm{s}$, can be obtained by direct integration of the pdf as in \Eqres{eq:r-mean} and \Eqrs{eq:r-variance} and applying the equivalence between the model parameters $(\alpha,\beta)$ and $(b,S,\kappa)$, \Eqre{eq:ab}.

DNS data of homogeneous RT mixing \cite{Livescu_07} indicates that as the two fluids mix, the density variance decays monotonically and approaches zero in the fully mixed state. \Eqre{eq:sdv} shows that $\rv{\varrho}\!\to\!0$ if and only if $\delta\!\to\!0$, independent of $S$. (As will be shown later, in the fully mixed limit $0\!\ne\!S\!\ne\!1$, hence the independence requirement.)

Since both terms on the right hand side of \Eqre{eq:dv} are always non-negative, the monotonicity of $\rv{\varrho}$ can only be ensured if
\begin{equation}
\delta < \frac{\rv{\varrho}}{\rmean{\varrho}(1-\rmean{\varrho})-\rv{\varrho}} = \frac{1-\theta}{\theta}.
\label{eq:constdelta}
\end{equation}
Thus in principle, \Eqre{eq:constdelta} could be used to constrain $\delta$ via
\begin{equation}
\delta = C_\delta\frac{\rv{\varrho}}{\rmean{\varrho}(1-\rmean{\varrho})-\rv{\varrho}} = C_\delta\frac{1-\theta}{\theta},
\label{eq:specdelta}
\end{equation}
with $0 \le C_\delta \le 1$. The above specification, however, may render the numerical method unstable even in the case of a non-fluctuating inverse time-scale, $b$, since $\kappa\!=\!\delta b$ drives the stochastic term in \Eqre{eq:dSDEt} and the denominator of \Eqre{eq:specdelta} is close to zero in the initial unmixed state. To eliminate this possibility we constrain $\delta$ via
\begin{equation}
\delta = C_\delta\frac{\rv{\varrho}}{\rmean{\varrho}(1-\rmean{\varrho})} = C_\delta(1-\theta),
\label{eq:delta}
\end{equation}
with $C_\delta$ to be determined by the model for the given application. Note that $C_\delta$, in general, need not be a constant, i.e.\ $C_\delta=C_\delta(t)$.

\Eqre{eq:delta} puts a stronger monotonicity constraint on the variance than \Eqre{eq:specdelta} would and assures that $\delta\to0$ if and only if $\rv{\varrho}\to0$, since $\rmean{\varrho}$ is bounded. Consequently, the constraint \Eqre{eq:delta} on $\delta$ establishes both physical and mathematical consistency of the variance evolution, since $\delta\to0 \iff \rv{\varrho}\to0$.

\Eqre{eq:dSDEt} has been constrained to a monotonic non-increasing variance, a fundamental requirement of mixing models in homogeneous flows.

\subsection{Mean: $\rmean{v}$}
Multiplying \Eqre{eq:dFP} by $v$ and integrating gives the evolution equation for the mean, $\rmean{v}$, of the beta-pdf as
\begin{equation}
\frac{1}{b}\frac{\partial\rmean{v}}{\partial t} = \left(\delta-\frac{S}{2}\right)\left(\rmean{v}^2+\rv{v}\right) + \left(\frac{1}{2}-\delta\right)\rmean{v}.\label{eq:vmean}
\end{equation}
Since the drift term in the SDE for $\varrho^*$, \Eqre{eq:dSDEt}, is linear, the drift in the equivalent SDE governing $v^*=1/\varrho^*$ is non-linear. As a consequence, in contrast to $\varrho$, the moment equations involving $v$ comprise an infinite hierarchy of non-closed system of differential equations.

\subsection{Covariance: $\rmean{\varrho\rf v\rf}$}
Multiplying \Eqre{eq:dFP} by $(\varrho-\rmean{\varrho})(v-\rmean{v})$ and integrating, or equivalently, employing the identity, $\rmean{\varrho\rf v\rf}=1-\rmean{\varrho}\cdot\rmean{v}$, and using \Eqres{eq:dmean} and \Eqrs{eq:vmean}, produce the evolution equation for the covariance $\rmean{\varrho\rf v\rf}$ as
\begin{equation}
\frac{1}{b}\frac{\partial\rmean{\varrho\rf v\rf}}{\partial t} = \left(\frac{S}{2}-\delta\right)\left(\rmean{v}+\frac{\rv{v}}{\rmean{v}}\right)\left(1-\rmean{\varrho\rf v\rf}\right) + \delta-\frac{S}{2}\rmean{v}.\label{eq:mrv}
\end{equation}

\subsection{Third moment: $\rs{\varrho}$}
Multiplying \Eqre{eq:dFP} by $(\varrho-\rmean{\varrho})^3$ and integrating result in the model evolution equation of the third moment of the beta-pdf
\ifthenelse{\boolean{JoT}}
{
\begin{align}
\frac{1}{3b}\frac{\partial\rs{\varrho}}{\partial t} & = \left[\delta(1-2\rmean{\varrho}) + \frac{1}{2}(S-\rmean{\varrho})\right]\rv{\varrho} - \left(\frac{1}{2}+\delta\right)\rs{\varrho},\nonumber\\
& = \left[\delta(1-2\rmean{\varrho}) + \frac{1}{2}(S-\rmean{\varrho})\right]\rmean{\varrho}(1-\rmean{\varrho})(1-\theta) - \left(\frac{1}{2}+\delta\right)\rs{\varrho},\label{eq:3m}
\end{align}
}
{
\begin{align}
\frac{1}{3b}\frac{\partial\rs{\varrho}}{\partial t} & = \left[\delta(1-2\rmean{\varrho}) + \frac{1}{2}(S-\rmean{\varrho})\right]\rv{\varrho} - \left(\frac{1}{2}+\delta\right)\rs{\varrho},\nonumber\\
& = \left[\delta(1-2\rmean{\varrho}) + \frac{1}{2}(S-\rmean{\varrho})\right]\rmean{\varrho}(1-\rmean{\varrho})(1-\theta)\nonumber\\
&\quad - \left(\frac{1}{2}+\delta\right)\rs{\varrho},\label{eq:3m}
\end{align}
}
whose stationary value, i.e.\ $\partial\rs{\varrho}/\partial t\!=\!0$, at any time gives the skewness as
\begin{equation}
\left.\frac{\rs{\varrho}}{\rv{\varrho}^{3/2}}\right|_\mathrm{s} = \frac{2\sqrt{\delta(1+\delta)}}{1+2\delta} \cdot \frac{1-2S}{\sqrt{S(1-S)}}.
\label{eq:ss}
\end{equation}
This means that for a symmetric distribution $S\!=\!1/2$, while $S\!<\!1/2$ and $S\!>\!1/2$ will result in positive and negative skewness, respectively.

\subsection{Fourth moment: $\rk{\varrho}$}
Multiplying \Eqre{eq:dFP} by $(\varrho-\rmean{\varrho})^4$ and integrating give the equation governing the fourth moment of the beta-pdf as
\ifthenelse{\boolean{JoT}}
{
\begin{align}
\frac{1}{2b}\frac{\partial\rk{\varrho}}{\partial t} & = 3\delta\rmean{\varrho}(1-\rmean{\varrho})\rv{\varrho} + \big[S-\rmean{\varrho}+3\delta(1-2\rmean{\varrho})\big]\rs{\varrho} - (1+3\delta)\rk{\varrho}\nonumber\\
& = 3\delta\rmean{\varrho}^2(1-\rmean{\varrho})^2(1-\theta) + \big[S-\rmean{\varrho}+3\delta(1-2\rmean{\varrho})\big]\rs{\varrho} - (1+3\delta)\rk{\varrho},\label{eq:4m}
\end{align}
}
{
\begin{align}
\frac{1}{2b}\frac{\partial\rk{\varrho}}{\partial t} & = 3\delta\rmean{\varrho}(1-\rmean{\varrho})\rv{\varrho} + \big[S-\rmean{\varrho}+3\delta(1-2\rmean{\varrho})\big]\rs{\varrho}\nonumber\\
&\quad - (1+3\delta)\rk{\varrho}\nonumber\\
& = 3\delta\rmean{\varrho}^2(1-\rmean{\varrho})^2(1-\theta) + \big[S-\rmean{\varrho}+3\delta(1-2\rmean{\varrho})\big]\rs{\varrho}\nonumber\\
&\quad - (1+3\delta)\rk{\varrho},\label{eq:4m}
\end{align}
}
which can be used to derive the stationary value to which the kurtosis will converge at any point in time
\begin{equation}
\left.\frac{\rk{\varrho}}{\rv{\varrho}^2}\right|_\mathrm{s} = (1+\delta)\left[\frac{3}{1+3\delta} + \frac{6\delta(1-2S)^2}{(1+3\delta)(1+2\delta)S(1-S)}\right].
\label{eq:sk}
\end{equation}

\subsection{The fully-mixed limit: $\delta\to0$}
Taking the limit $\delta\to0$ of the stationary skewness and kurtosis in \Eqres{eq:ss} and \Eqrs{eq:sk} shows that by specifying $\delta$ via the monotonically decreasing function of \Eqre{eq:delta}, at $t\to\infty$ the skewness will vanish, while the kurtosis will approach the Gaussian value of 3, independent of $S$.

The above development is consistent with Girimaji's \cite{Girimaji_91} analysis of the beta-pdf in the limit of small variance: by satisfying the monotonicity constraint on $\delta$, \Eqre{eq:delta}, the SDE \Eqrs{eq:dSDEt} approximates a clipped Gaussian as $t\to\infty$.

To summarize, in the fully-mixed limit, $\delta\to0$, the stationary values of the first four moments, obtained from \Eqres{eq:sdmean}, \Eqrs{eq:sdv}, \Eqrs{eq:ss} and \Eqrs{eq:sk} will be
\begin{align}
\rmean{\varrho}^\mathrm{fm}_\mathrm{s} & = S,\\
\rv{\varrho}^\mathrm{fm}_\mathrm{s} & = 0,\\
\left.\frac{\rs{\varrho}}{\rv{\varrho}^{3/2}}\right|^\mathrm{fm}_\mathrm{s} & = 0,\\
\left.\frac{\rk{\varrho}}{\rv{\varrho}^2}\right|^\mathrm{fm}_\mathrm{s} & = 3.
\end{align}
These equations characterise the asymptotic shape of the pdf at $t\to\infty$.

\subsection{The no-mix limit: $\delta\to\infty$}
Mathematically, the no-mix limit corresponds to $\delta\to\infty$ or equivalently $\theta\to0$. Strictly speaking, \Eqre{eq:beta} is undefined for $\delta=\infty$, i.e.\ $\alpha=\beta=0$, nevertheless it is useful to examine $\delta\to\infty$ as an asymptotic limit. This results in the following stationary formulas for the first four moments, obtained from \Eqres{eq:sdmean}, \Eqrs{eq:sdv}, \Eqrs{eq:ss} and \Eqrs{eq:sk},
\begin{align}
\rmean{\varrho}^\mathrm{nm}_\mathrm{s} & = S,\label{eq:nm1}\\
\rv{\varrho}^\mathrm{nm}_\mathrm{s} & = S(1-S),\label{eq:nm2}\\
\left.\frac{\rs{\varrho}}{\rv{\varrho}^{3/2}}\right|^\mathrm{nm}_\mathrm{s} & = \frac{1-2S}{\sqrt{S(1-S)}},\label{eq:nm3}\\
\left.\frac{\rk{\varrho}}{\rv{\varrho}^2}\right|^\mathrm{nm}_\mathrm{s} & = \frac{1}{S(1-S)} - 3.\label{eq:nm4}
\end{align}
These equations characterise the shape of the pdf independent of molecular diffusion. They show the effect of mixing asymmetry on the statistics and that even if mixing is not allowed, the distribution is capable of representing asymmetry (and skewness) by specifying the time evolution of the single model parameter $S(t)$.

\subsection{Summary}
The equations governing the statistics of the SDE \Eqrs{eq:dSDEt} are summarized in Table \ref{tab:moment_eqs}. It is emphasized that these moment equations are 1) independent of any physics and 2) a precise result of pure mathematical nature: they represent the governing equations of the moments of a non-stationary beta distribution.
\begin{table}
\caption{The equations governing the statistics of the non-stationary beta-pdf, derived from the SDE \Eqrs{eq:dSDEt} with time-varying coefficients, $b(t)$, $S(t)$ and $\delta(t)=\kappa(t)/b(t)$.}
\hrule\addvspace{2pt}\hrule
\begin{align*}
\frac{\partial\rmean{\varrho}}{\partial t} & = \frac{b}{2}(S-\rmean{\varrho})\\
\frac{1}{b}\frac{\partial\rv{\varrho}}{\partial t} & = \delta\rmean{\varrho}(1-\rmean{\varrho}) - (1+\delta)\rv{\varrho}\\
\frac{1}{b}\frac{\partial\rmean{v}}{\partial t} & = \left(\delta-\frac{S}{2}\right)\left(\rmean{v}^2+\rv{v}\right) + \left(\frac{1}{2}-\delta\right)\rmean{v}\\
\frac{1}{b}\frac{\partial\rmean{\varrho\rf v\rf}}{\partial t} & = \left(\frac{S}{2}-\delta\right)\left(\rmean{v}+\frac{\rv{v}}{\rmean{v}}\right)\left(1-\rmean{\varrho\rf v\rf}\right) + \delta-\frac{S}{2}\rmean{v}\\
\frac{1}{3b}\frac{\partial\rs{\varrho}}{\partial t} & = \left[\delta(1-2\rmean{\varrho}) + \frac{1}{2}(S-\rmean{\varrho})\right]\rv{\varrho} - \left(\frac{1}{2}+\delta\right)\rs{\varrho}\\
\frac{1}{2b}\frac{\partial\rk{\varrho}}{\partial t} & = 3\delta\rmean{\varrho}(1-\rmean{\varrho})\rv{\varrho} + \big[S-\rmean{\varrho}+3\delta(1-2\rmean{\varrho})\big]\rs{\varrho} - (1+3\delta)\rk{\varrho}
\end{align*}
\hrule\addvspace{2pt}\hrule
\label{tab:moment_eqs}
\end{table}

The constraints that must be satisfied by the coefficients, $b(t)$, $S(t)$ and $\kappa(t)$, for mathematical and physical realizability of a material mixing model in homogeneous flows, governed by \Eqre{eq:dSDEt}, are
\begin{enumerate}
\item Positivity of the inverse time-scale, $b(t)>0$,
\item Boundedness of $0 < S(t) < 1$, and,
\item To ensure a non-increasing variance, \Eqre{eq:delta}, with $0 \le C_\delta \le 1$.
\end{enumerate}

The no-mix (or high-\textit{Sc}) limit is obtained if $\kappa(t) \gg b(t)$ at all times.

The above development provided necessary conditions for mathematical and physical consistency of \Eqre{eq:dSDEt} for its use as a material mixing model for passive scalars. In the next section the model will be shown to satisfy conservation of mass which establishes sufficient conditions for its application as a model for the stochastic fluid density.

\section{Consistency with mass conservation}
\label{sec:mass_consistency}
Up to this point, the development has been purely mathematical in nature; we have made no restrictions regarding what physical quantity $\varrho$ can represent, only that it is a beta-distributed scalar. By constructing the time-evolution of the model coefficients $\delta=\kappa/b$ to satisfy \Eqre{eq:delta}, the governing SDE \Eqrs{eq:dSDEt} has been confined to a temporal evolution of the particle property $\varrho^*$ whose variance, $\rv{\varrho}$, cannot increase, a fundamental requirement of material mixing models in statistically homogeneous flows. The extension to inhomogeneous flows, which introduces no new small scale terms related to molecular mixing, is made subsequently.

In the following, physics is introduced by assuming that $\varrho^*$ represents the fluid density, $\varrho$. The continuum form of mass conservation is used to derive the equations for the moments of the density distribution. Comparisons between the moment equations from the SDE (a mathematical model for the evolution of a non-stationary beta-pdf) and the moment equations from mass conservation (a mathematical model expressing the physical conservation principles) is required in order to relate the SDE parameters, $b$, $S$ and $\kappa$, to physical processes of molecular mixing.

\subsection{Moment equations from conservation of mass}
The starting point for the exact equations is the conservation of mass along an instantaneous Lagrangian path for the density, $\varrho$, and specific volume, $v=1/\varrho$, respectively,
\begin{align}
\ld{\varrho}=-\varrho\vd \qquad \mathrm{and} \qquad \ld{v}=v\vd,\label{eq:mass}
\end{align}
with the dilatation $d=v_{i,i}$ and the Lagrangian derivative
\begin{equation}
\ld{} \equiv \frac{\partial}{\partial t} + v_k\frac{\partial}{\partial x_k}.\label{eq:ld}
\end{equation}

As a matter of clarity and convenience, the Lagrangian notation will be used, as it reduces the number of terms in the equations. (In a subsequent section, \ref{sec:inhom}.\ \emph{Extension to inhomogeneous flows}, Eulerian notation is used.) In joint pdf methods containing the velocity, terms originating from the physical process of advection appear in closed form; these represent mean and turbulent transport and production/destruction. For simplicity we choose to incorporate these in the Lagrangian derivative. It is worth emphasizing, that this also means that the equations for statistics will represent the \emph{rate of change along instantaneous Lagrangian paths}, as defined by \Eqre{eq:ld}. In the following, we assume the existence of a velocity pdf model. In joint pdf methods for a set of scalars, where the full velocity pdf is unavailable, turbulent transport \cite{Fox_03} and, as will be shown, the mass flux require closure assumptions.

If the particle position, $x_i^*$, is governed by
\begin{equation}
\mathrm{d}x_i^* = v_i^*\mathrm{d}t,
\label{eq:x}
\end{equation}
the model FPE governing the Eulerian joint pdf of density and velocity, $f(\varrho,\bv{v};\bv{x},t)$, can be stated as
\begin{align}
\frac{\partial f}{\partial t} + v_i\frac{\partial f}{\partial x_i} = & - \frac{\partial}{\partial\varrho}\left[\frac{b}{2}(S-\varrho)f\right] + \frac{1}{2}\frac{\partial^2}{\partial\varrho^2}\big[\kappa\varrho(1-\varrho)f\big]\nonumber\\
& + \textrm{velocity model terms.}
\label{eq:FP}
\end{align}

\textbf{The mean density equation: $\rmean{\varrho}$.} The exact equation in homogeneous flows, derived from \Eqre{eq:mass}, is
\begin{equation}
\ld{\rmean{\varrho}} = -\rmean{\varrho\rf\vdrf},\label{eq:ed1}
\end{equation}
where $\ivdrf=v\rf_{i,i}$. The evolution equation for the mean density along a Lagrangian path, according to the joint pdf model, \Eqre{eq:FP}, is
\begin{equation}
\left.\ld{\rmean{\varrho}}\right|_\mathrm{sde} = \frac{b}{2}(S-\rmean{\varrho}).\label{eq:mmd}
\end{equation}
The correlation of the fluctuating density and velocity divergence is modelled by the mean of the drift term in \Eqre{eq:dSDEt} as
\begin{equation}
-\rmean{\varrho\rf\vdrf} = \frac{b}{2}(S-\rmean{\varrho}).\label{eq:im1}
\end{equation}
The SDE \Eqrs{eq:dSDEt}, representing the fluid density, is consistent with conservation of mass in the mean if the above holds.

\textbf{The density variance equation: $\rv{\varrho}$.} For a statistically homogeneous flow the exact and the pdf-model density variance equations are
\begin{align}
\ld{\rv{\varrho}} & = -2\rmean{\varrho}\!\cdot\!\rmean{\varrho\rf\vdrf} - 2\rmean{\varrho\rf^2\vdrf},\label{eq:ed2}\\
\left.\ld{\rv{\varrho}}\right|_\mathrm{sde} & = \kappa\rmean{\varrho}(1-\rmean{\varrho}) - (b+\kappa)\rv{\varrho}.\label{eq:mvd}
\end{align}
Thus the mixing terms in the exact equation are jointly represented by the terms in the model equation as
\begin{equation}
- 2\rmean{\varrho}\!\cdot\!\rmean{\varrho\rf\vdrf} - 2\rmean{\varrho\rf^2\vdrf} = \kappa\rmean{\varrho}(1-\rmean{\varrho}) - (b+\kappa)\rv{\varrho}.\label{eq:im2}
\end{equation}

\textbf{The mean specific-volume equation: $\rmean{v}$.} The exact and model equations governing the mean specific volume are, respectively,
\begin{align}
\ld{\rmean{v}} & = \rmean{v\rf\vdrf},\label{eq:es1}\\
\left.\ld{\rmean{v}}\right|_\mathrm{sde} & = \left(\kappa-\frac{b}{2}S\right)\left(\rmean{v}^2+\rv{v}\right) + \left(\frac{b}{2}-\kappa\right)\rmean{v},\label{eq:mmsv}
\end{align}
indicating the following relation between the dissipation of $\rmean{v}$ and the SDE parameters:
\begin{equation}
\rmean{v\rf\vdrf} =  \left(\kappa-\frac{b}{2}S\right)\left(\rmean{v}^2+\rv{v}\right) + \left(\frac{b}{2}-\kappa\right)\rmean{v}.\label{eq:mmv}
\end{equation}

\textbf{The density-specific-volume covariance equation: $\rmean{\varrho\rf v\rf}$.} In variable-density turbulence the quantity $\rmean{\varrho\rf v\rf}$ plays a primary role in the production of the mass flux which, in the presence of a mean pressure gradient, drives the turbulence \cite{Livescu_09}. Note that $1 - \rmean{\varrho}\!\cdot\!\rmean{v} = \rmean{\varrho\rf v\rf}$.

The exact and pdf-model evolution equations for the density-specific-volume covariance are
\begin{align}
\ld{\rmean{\varrho\rf v\rf}} & = \rmean{v}\!\cdot\!\rmean{\varrho\rf\vdrf} - \rmean{\varrho}\!\cdot\!\rmean{v\rf\vdrf},\label{eq:edsv}\\
\left.\ld{\rmean{\varrho\rf v\rf}}\right|_\mathrm{sde} & = \left(\frac{S}{2}-\delta\right)\bigg(\rmean{v}+\frac{\rv{v}}{\rmean{v}}\bigg)\left(1-\rmean{\varrho\rf v\rf}\right) + \delta-\frac{S}{2}\rmean{v}.\label{eq:mrvi}
\end{align}
As above, a comparison of the right hand sides of \Eqres{eq:edsv} and \Eqrs{eq:mrvi} relates the physical mixing processes to the SDE parameters.

\textbf{The third density moment equation: $\rs{\varrho}$.} The exact and the pdf-model governing equations for the third moment are
\begin{align}
\frac{1}{3}\ld{\rs{\varrho}} & = \rv{\varrho}\!\cdot\!\rmean{\varrho\rf\vdrf} - \rmean{\varrho}\!\cdot\!\rmean{\varrho\rf^2\vdrf} - \rmean{\varrho\rf^3\vdrf},\label{eq:ed3}\\
\frac{1}{3}\left.\ld{\rs{\varrho}}\right|_\mathrm{sde} & = \left[\kappa(1-2\rmean{\varrho}) + \frac{b}{2}(S-\rmean{\varrho})\right]\rv{\varrho} - \left(\frac{b}{2}+\kappa\right)\rs{\varrho}.\label{eq:dms}
\end{align}
Comparing the right hand sides shows how the SDE parameters are related to the mixing processes:
\ifthenelse{\boolean{JoT}}
{
\begin{equation}
\rv{\varrho}\!\cdot\!\rmean{\varrho\rf\vdrf} - \rmean{\varrho}\!\cdot\!\rmean{\varrho\rf^2\vdrf} - \rmean{\varrho\rf^3\vdrf} = \left[\kappa(1-2\rmean{\varrho}) + \frac{b}{2}(S-\rmean{\varrho})\right]\rv{\varrho} - \left(\frac{b}{2}+\kappa\right)\rs{\varrho}.\label{eq:im3}
\end{equation}
}
{
\begin{align}
\begin{split}
\rv{\varrho}\!&\cdot\!\rmean{\varrho\rf\vdrf} - \rmean{\varrho}\!\cdot\!\rmean{\varrho\rf^2\vdrf} - \rmean{\varrho\rf^3\vdrf} =\\
& = \left[\kappa(1-2\rmean{\varrho}) + \frac{b}{2}(S-\rmean{\varrho})\right]\rv{\varrho} - \left(\frac{b}{2}+\kappa\right)\rs{\varrho}.\label{eq:im3}
\end{split}
\end{align}
}

\textbf{The fourth density moment equation: $\rk{\varrho}$.} The exact and modelled fourth moment equations are, respectively,
\ifthenelse{\boolean{JoT}}
{
\begin{align}
\frac{1}{4}\ld{\rk{\varrho}} & = \rs{\varrho}\!\cdot\!\rmean{\varrho\rf\vdrf} - \rmean{\varrho}\!\cdot\!\rmean{\varrho\rf^3\vdrf} - \rmean{\varrho\rf^4\vdrf},\label{eq:ed4}\\
\frac{1}{4}\left.\ld{\rk{\varrho}}\right|_\mathrm{sde} & = \frac{3}{2}\kappa\rmean{\varrho}(1-\rmean{\varrho})\rv{\varrho} + \frac{1}{2}\big[b(S-\rmean{\varrho}) + 3\kappa(1-2\rmean{\varrho})\big]\rs{\varrho} - \frac{1}{2}(b+3\kappa)\rk{\varrho}.\label{eq:dmk}
\end{align}
}
{
\begin{align}
\frac{1}{4}\ld{\rk{\varrho}} & = \rs{\varrho}\!\cdot\!\rmean{\varrho\rf\vdrf} - \rmean{\varrho}\!\cdot\!\rmean{\varrho\rf^3\vdrf} - \rmean{\varrho\rf^4\vdrf},\label{eq:ed4}\\
\frac{1}{4}\left.\ld{\rk{\varrho}}\right|_\mathrm{sde} & = \frac{3}{2}\kappa\rmean{\varrho}(1-\rmean{\varrho})\rv{\varrho} - \frac{1}{2}(b+3\kappa)\rk{\varrho}\nonumber\\
&\quad + \frac{1}{2}\big[b(S-\rmean{\varrho}) + 3\kappa(1-2\rmean{\varrho})\big]\rs{\varrho}.\label{eq:dmk}
\end{align}
}
Comparing the right hand sides shows how the SDE parameters are related to the mixing processes:
\ifthenelse{\boolean{JoT}}
{
\begin{align}
\rs{\varrho}\!\cdot\!\rmean{\varrho\rf\vdrf} & - \rmean{\varrho}\!\cdot\!\rmean{\varrho\rf^3\vdrf} - \rmean{\varrho\rf^4\vdrf} = \nonumber\\
& = \frac{3}{2}\kappa\rmean{\varrho}(1-\rmean{\varrho})\rv{\varrho} + \frac{1}{2}\big[b(S-\rmean{\varrho}) + 3\kappa(1-2\rmean{\varrho})\big]\rs{\varrho} - \frac{1}{2}(b+3\kappa)\rk{\varrho}.\label{eq:im4}
\end{align}
}
{
\begin{align}
\rs{\varrho}\!\cdot\!\rmean{\varrho\rf\vdrf} & - \rmean{\varrho}\!\cdot\!\rmean{\varrho\rf^3\vdrf} - \rmean{\varrho\rf^4\vdrf} = \nonumber\\
& = \frac{3}{2}\kappa\rmean{\varrho}(1-\rmean{\varrho})\rv{\varrho} - \frac{1}{2}(b+3\kappa)\rk{\varrho}\nonumber\\
&\quad + \frac{1}{2}\big[b(S-\rmean{\varrho}) + 3\kappa(1-2\rmean{\varrho})\big]\rs{\varrho}.\label{eq:im4}
\end{align}
}

The above development, relating the SDE parameters to physical processes, are a rigorous mathematical consequence of one assumption: the fluid mass density in a homogeneous flow is beta-distributed.

\subsection{Closure relations for $\rmean{\varrho\rf^n\ivdrf}$, $n\ge1$}
\Eqres{eq:im1}, \Eqrs{eq:im2}, \Eqrs{eq:im3} and \Eqrs{eq:im4} indicate that assuming a beta-pdf for the fluid density, the moment equations are inter-dependent and imply a series of relations for the density-dilatation statistics $\rmean{\varrho\rf^{n}\ivdrf}$, $n\ge1$. These relations are now explicitly detailed.

\Eqre{eq:im1} indicates that
\begin{equation}
\rmean{\varrho\rf\vdrf} = - \frac{b}{2}(S-\rmean{\varrho}),\label{eq:mi1}
\end{equation}
which, when substituted into \Eqre{eq:im2}, implies
\begin{equation}
\rmean{\varrho\rf^2\vdrf} = \frac{1}{2}\left[b\rmean{\varrho}(S-\rmean{\varrho}) - \kappa\rmean{\varrho}(1-\rmean{\varrho}) + (b+\kappa)\rv{\varrho}\right].\label{eq:mi2}
\end{equation}
Then substituting both \Eqres{eq:mi1} and \Eqrs{eq:mi2} into \Eqre{eq:im3} results in
\begin{align}
\rmean{\varrho\rf^3\vdrf} & = \big[\kappa(1-\rmean{\varrho}) - b(S-\rmean{\varrho})\big]\frac{\rmean{\varrho}^2}{2} + \left(\frac{b}{2}+\kappa\right)\rs{\varrho}\nonumber\\
& \quad - \left[b(S-\rmean{\varrho})+\frac{\rmean{\varrho}}{2}(b+\kappa)+\kappa(1-2\rmean{\varrho})\right]\rv{\varrho}.\label{eq:mi3}
\end{align}

Eqs.\ (\ref{eq:mi1}--\ref{eq:mi3}), provide a series of relations for the covariances of the form $\rmean{\varrho\rf^n\ivdrf}$, establishing a number of relations between various mixing processes and the SDE parameters.

\textbf{Moment and SDE relations in the no-mix limit.} It is useful to investigate the closures, $\rmean{\varrho\rf^n\ivdrf}$, Eqs.\ (\ref{eq:mi1}--\ref{eq:mi3}), in a perturbed state about the no-mix limit, $\theta=\varepsilon_\theta$, where $0<\varepsilon_\theta\ll1$. This results in 
\begin{align}
\frac{2}{b}\rmean{\varrho\rf\vdrf}^\mathrm{nm}_\varepsilon & = -(S-\rmean{\varrho}),\label{eq:mi1-nm}\\
\frac{2}{b}\frac{\rmean{\varrho\rf^2\vdrf}^\mathrm{nm}_\varepsilon}{\rmean{\varrho}(1-\rmean{\varrho})} & = \frac{S-\rmean{\varrho}}{1-\rmean{\varrho}} - \delta + (1+\delta)(1-\varepsilon_\theta),\label{eq:mi2-nm}\\
\frac{2}{b}\frac{\rmean{\varrho\rf^3\vdrf}^\mathrm{nm}_\varepsilon}{\rmean{\varrho}(1-\rmean{\varrho})} & = \left(\delta-\frac{S-\rmean{\varrho}}{1-\rmean{\varrho}}\right)\rmean{\varrho} + (1+2\delta)(1-2\rmean{\varrho})(1-\varepsilon_\theta)\nonumber\\
&\ifthenelse{\boolean{JoT}}{\quad}{} -\big[2(S-\rmean{\varrho}) + \rmean{\varrho}(1+\delta) + 2\delta(1-2\rmean{\varrho})\big](1-\varepsilon_\theta).\label{eq:mi3-nm}
\end{align}
If $\delta\to\infty$ and $\varepsilon_\theta\to0$ at the same rate, Eqs.\ (\ref{eq:mi1-nm}--\ref{eq:mi3-nm}) result in well-defined (finite) expressions for all $\rmean{\varrho\rf^n\ivdrf}$ in the no-mix state.

\subsection{Connecting $b$, $S$ and $\kappa$ to physical processes}
Applying the results of setting the right hand side of the moment equations that come from continuity to the right hand sides of the moment equations that come from the SDE allows one to express $b$, $S$ and $\kappa$ in terms of physical mixing processes.

Solving the relations for $\rmean{\varrho\rf^{n}\ivdrf}$, $n\le3$, Eqs.\ (\ref{eq:mi1}--\ref{eq:mi3}), for $b$, $S$ and $\kappa$ results in
\begin{align}
b & = \frac{1}{F}\Lambda_1 + 2\frac{1-F}{F}\Lambda_3,\label{eq:bphys}\\
\kappa & = \frac{1}{F}\frac{\psi}{1-\psi}\Lambda_2,\label{eq:kphys}\\
S & = \rmean{\varrho} - \frac{2}{b}\rmean{\varrho\rf\vdrf},\label{eq:Sphys}
\end{align}
with
\begin{align}
\Lambda_1 & = \frac{2\rmean{\varrho}\!\cdot\!\rmean{\varrho\rf\vdrf} + 2\rmean{\varrho\rf^2\vdrf}}{\rv{\varrho}},\label{eq:L1}\\
\Lambda_2 & = \bigg(2\frac{\rv{\varrho}}{\rs{\varrho}} + \frac{\rmean{\varrho}}{\rv{\varrho}}\bigg)\rmean{\varrho\rf\vdrf} + \left(\frac{1}{\rv{\varrho}} - \frac{\rmean{\varrho}}{\rs{\varrho}}\right)\rmean{\varrho\rf^2\vdrf} - \frac{\rmean{\varrho\rf^3\vdrf}}{\rs{\varrho}},\label{eq:L2}\\
\Lambda_3 & = 2\rv{\varrho}\!\cdot\!\rmean{\varrho\rf\vdrf} - \rmean{\varrho}\!\cdot\!\rmean{\varrho\rf^2\vdrf} - \rmean{\varrho\rf^3\vdrf},\label{eq:L3}\\
F & = 1 - \frac{1}{2}\frac{\theta\psi}{(1-\theta)(1-\psi)},
\end{align}
and
\begin{equation}
\psi = \frac{\rs{\varrho}}{(1-2\rmean{\varrho})\rv{\varrho}} = \frac{\rv{\varrho}^{1/2}}{1-2\rmean{\varrho}}\cdot\frac{\rs{\varrho}}{\rv{\varrho}^{3/2}}.\label{eq:psi}
\end{equation}

Eqs.\ (\ref{eq:bphys}--\ref{eq:Sphys}) show how the first three density-velocity-dilatation covariances together with the first three density moments determine the three SDE coefficients in \Eqre{eq:dSDEt} or, equivalently, the two parameters of the beta-pdf, \Eqre{eq:ab}.

The unit of $\Lambda_1$ and $\Lambda_2$ is that of $\ivdrf$, the mixing rate, indicating their effect on the mixing state, while $\Lambda_3$ corresponds to asymmetry.

Dividing \Eqres{eq:kphys} and \Eqrs{eq:bphys} yields the useful result
\begin{equation}
\delta = \frac{1-\theta}{\theta} + \frac{S-\rmean{\varrho}}{\theta\rmean{\varrho}(1-\rmean{\varrho})}\left(\rmean{\varrho}+\frac{\rmean{\varrho\rf^2\vdrf}}{\rmean{\varrho\rf\vdrf}}\right)\label{eq:dphys}
\end{equation}
\Eqre{eq:dphys} is partitioned into two parts, responsible for the symmetric and non-symmetric behaviour of the pdf, respectively. The first term (from $\Lambda_1$) is the effect of molecular mixing, c.f.\ \Eqre{eq:constdelta}, and the second (from $\Lambda_3$) is the mixing asymmetry. Note that \Eqre{eq:constdelta} is a model constraint, while \Eqre{eq:dphys} is a relationship of $\delta$ to physics if the underlying pdf is beta.

The parameters, $b$, $S$ and $\kappa$, of the beta-SDE, representing the fluid mass density, have been related to physical mixing processes.

\textbf{The SDE parameters in the no-mix limit.} It is insightful to investigate the relations for $b$, $S$ and $\kappa$, Eqs.\ (\ref{eq:bphys}--\ref{eq:Sphys}) in the no-mix (or high-\textit{Sc}) limit.

Note that $\Phi=1-\psi$ is another mix-metric (similar to $\theta$) with $\Phi=0$ in the no-mix and $\Phi=1$ in the fully mixed states. \Eqre{eq:psi} shows that $\psi$ is only defined for a non-symmetric distribution with finite skewness (see also Appendix \ref{app:sym}), which can be easily seen in the no-mix limit:
\begin{equation}
\psi^\mathrm{nm} = \frac{\sqrt{S(1-S)}}{1-2S}\cdot\frac{1-2S}{\sqrt{S(1-S)}} = 1
\ifthenelse{\boolean{JoT}}{\qquad}{\quad} \mathrm{for} \ifthenelse{\boolean{JoT}}{\qquad}{\quad} S \ne 0, \frac{1}{2}, 1.
\end{equation}

Denoting small departures from the no-mix state by $\varepsilon_\theta>0$ and $\varepsilon_\psi>0$, i.e.\ $\theta=\varepsilon_\theta$ and $\psi=1-\varepsilon_\psi$, the SDE coefficients in the perturbed state are given by
\begin{align}
b^\mathrm{nm}_\varepsilon & = \frac{2\varepsilon_\psi(1-\varepsilon_\theta)}{2\varepsilon_\psi(1-\varepsilon_\theta) - \varepsilon_\theta(1-\varepsilon_\psi)}\Lambda_1 + \frac{\varepsilon_\theta(1-\varepsilon_\psi)}{\varepsilon_\psi(1-\varepsilon_\theta)}\Lambda_3,\label{eq:bnm}\\
\kappa^\mathrm{nm}_\varepsilon & = \frac{2(1-\varepsilon_\theta)(1-\varepsilon_\psi)}{2\varepsilon_\psi(1-\varepsilon_\theta) - \varepsilon_\theta(1-\varepsilon_\psi)}\Lambda_2,\label{eq:knm}\\
S^\mathrm{nm}_\varepsilon & = \rmean{\varrho} - \frac{2}{b^\mathrm{nm}_\varepsilon}\rmean{\varrho\rf\vdrf}.\label{eq:Snm}
\end{align}
Further simplifying with $\varepsilon_\psi = \varepsilon_\theta/c = \varepsilon$ with the positive constant $0<c<\infty$, \Eqres{eq:bnm} and \Eqrs{eq:knm} become
\begin{align}
b^\mathrm{nm}_\varepsilon & = \frac{2-2c\varepsilon}{2-c-c\varepsilon}\Lambda_1 + \frac{c-c\varepsilon}{1-c\varepsilon}\Lambda_3,\label{eq:bnm2}\\
\kappa^\mathrm{nm}_\varepsilon & = \frac{2-2(1+c)\varepsilon+2c\varepsilon^2}{(2-c)\varepsilon-c\varepsilon^2}\Lambda_2.\label{eq:knm2}
\end{align}
\Eqre{eq:bnm2} indicates that $b(t)$, the model mixing rate, starts from a well-defined finite value in the unmixed state, $\varepsilon\to0$,
\begin{equation}
b^\mathrm{nm}_{\varepsilon\to0} = \frac{2}{2-c}\Lambda_1 + c\Lambda_3,\label{eq:bnm3}
\end{equation}
while repeatedly applying L'H\^opital's rule to \Eqre{eq:knm2} leads to
\begin{equation}
\kappa^\mathrm{nm}_{\varepsilon\to0} = -2\Lambda_2,\label{eq:knm3}
\end{equation}
and thus
\begin{equation}
\delta^\mathrm{nm}_{\varepsilon\to0} = -\frac{2(2-c)\Lambda_2}{2\Lambda_1+c(2-c)\Lambda_3}.
\end{equation}
Assuming the same perturbation in both mix-metrics, i.e.\ $c=1$, results in
\begin{equation}
\delta^\mathrm{nm}_{\varepsilon\to0,c=1} = -\frac{2\Lambda_2}{2\Lambda_1+\Lambda_3}.\label{eq:dnm}
\end{equation}

The SDE parameters have been related to physical mixing processes in the perturbed no-mix limit.

\subsection{Summary}
This section detailed the consequences of representing the fluid mass density by a beta distribution in homogeneous non-stationary mixing flows. In summary, 

\begin{enumerate}
\item The time-inhomogeneous governing equation \Eqrs{eq:dSDEt} representing the density is consistent with conservation of mass.

\item A series of relations for the correlations between the density and the velocity dilatation of the form, $\rmean{\varrho\rf^{n}\ivdrf}$, $n\ge1$, has been obtained.

\item The physical meaning of the three model coefficients, $b(t)$, $S(t)$ and $\kappa(t)$, have been explicitly related to the mixing physics, as reflected in the density moments and density-dilatation covariances.

\item If one were to design a moment closure for the first few moments of the pdf, one can now relate the mixing statistics in the various moment equations to each other by one consistency principle: models for the mixing processes in the different moment equation are related to each other in a unique self-consistent way if the underlying pdf is beta.
\end{enumerate}

\section{Extension to inhomogeneous flows}
\label{sec:inhom}
The SDE \Eqrs{eq:dSDEt} is now extended to inhomogeneous flows. Beside micro-mixing, this allows the SDE to represent different macro-mixed states \cite{Fox_03}.

\subsection{Exact inhomogeneous equations}
The governing moment equations, derived from continuity, are given in both Lagrangian and Eulerian frameworks, establishing a correspondence between the pdf approach and moment closures.

The equations governing the density and specific volume statistics, derived from \Eqre{eq:mass}, in inhomogeneous flows are
{\allowdisplaybreaks
\begin{align}
\ld{\rmean{\varrho}} & = -\rmean{\varrho}\!\cdot\!\rmean{\vd} - \rmean{\varrho\rf\vdrf},\label{eq:ed1i}\\
\frac{1}{2}\ld{\rv{\varrho}} & = -\rv{\varrho}\!\cdot\!\rmean{\vd} - \rmean{\varrho}\!\cdot\!\rmean{\varrho\rf\vdrf} - \rmean{\varrho\rf^2\vdrf},\label{eq:ed2i}\\
\ld{\rmean{v}} & = \rmean{v}\!\cdot\!\rmean{\vd} + \rmean{v\rf\vdrf},\label{eq:es1i}\\
\ld{\rmean{\varrho\rf v\rf}} & = \rmean{v}\!\cdot\!\rmean{\varrho\rf\vdrf} - \rmean{\varrho}\!\cdot\!\rmean{v\rf\vdrf},\label{eq:edsvi}\\
\frac{1}{3}\ld{\rs{\varrho}} & = -\rs{\varrho}\!\cdot\!\rmean{\vd} + \rv{\varrho}\!\cdot\!\rmean{\varrho\rf\vdrf} - \rmean{\varrho}\!\cdot\!\rmean{\varrho\rf^2\vdrf} - \rmean{\varrho\rf^3\vdrf},\label{eq:ed3i}\\
\frac{1}{4}\ld{\rk{\varrho}} & = -\rk{\varrho}\!\cdot\!\rmean{\vd} + \rs{\varrho}\!\cdot\!\rmean{\varrho\rf\vdrf} - \rmean{\varrho}\!\cdot\!\rmean{\varrho\rf^3\vdrf} - \rmean{\varrho\rf^4\vdrf}.\label{eq:ed4i}
\end{align}}%
These equations are along an instantaneous Lagrangian path. The Lagrangian derivative, \Eqre{eq:ld}, can be used to write \Eqre{eq:ed1i} in equivalent forms more traditional in moment closures as
 \begin{align}
\frac{\partial\rmean{\varrho}}{\partial t} + \rmean{v}_i\!\cdot\!\rmean{\varrho},_i + \rmean{v_i\rf\varrho\rf\!,_i} & = -\rmean{\varrho}\!\cdot\!\rmean{\vd} - \rmean{\varrho\rf\vdrf},\label{eq:emdtr}\\
\frac{\partial\rmean{\varrho}}{\partial t} + \frac{\partial\rmean{\varrho}\fmean{v_i}}{\partial x_i} & = 0 \qquad \mathrm{with} \qquad \fmean{v_i}=\frac{\rmean{\varrho v_i}}{\rmean{\varrho}},\label{eq:emdt}
 \end{align}
where the Favre average is denoted by $\fmean{\thinspace\cdot\thinspace}$. Similarly, the exact variance equation \Eqrs{eq:ed2i} in equivalent Eulerian form is
\ifthenelse{\boolean{JoT}}
{
\begin{equation}
\frac{\partial\rv{\varrho}}{\partial t} + \rmean{v}_i\frac{\partial\rv{\varrho}}{\partial x_i} - 2\rmean{\varrho}\!\cdot\!\rmean{v_i\ff}\!\cdot\!\rmean{\varrho},_i + \frac{\partial\rmean{\varrho\rf^2v_i\rf}}{\partial x_i} = -2\rv{\varrho}\!\cdot\!\rmean{d} - 2\rmean{\varrho}\!\cdot\!\rmean{\varrho\rf\vdrf} - \rmean{\varrho\rf^2\vdrf},\label{eq:evdt}
\end{equation}
}
{
\begin{align}
\frac{\partial\rv{\varrho}}{\partial t} + \rmean{v}_i\frac{\partial\rv{\varrho}}{\partial x_i} & - 2\rmean{\varrho}\!\cdot\!\rmean{v_i\ff}\!\cdot\!\rmean{\varrho},_i + \frac{\partial\rmean{\varrho\rf^2v_i\rf}}{\partial x_i} =\nonumber\\
&\quad = -2\rv{\varrho}\!\cdot\!\rmean{d} - 2\rmean{\varrho}\!\cdot\!\rmean{\varrho\rf\vdrf} - \rmean{\varrho\rf^2\vdrf},\label{eq:evdt}
\end{align}
}
where the velocity fluctuation about the Favre average is $v_i\ff = v_i-\ifmean{v_i}$. Expanding the advection term in \Eqre{eq:es1i} leads to
\begin{equation}
\frac{\partial\rmean{v}}{\partial t} + (\rmean{v}\!\cdot\!\rmean{v}_i + \rmean{v\rf v_i\rf}),_i = 2\rmean{v}\!\cdot\!\rmean{d} + 2\rmean{v\rf\ivdrf}.\label{eq:V}
\end{equation}
The equation for the density-specific-volume covariance in the Eulerian framework is
\ifthenelse{\boolean{JoT}}
{
\begin{equation}
\frac{\partial\rmean{\varrho\rf v\rf}}{\partial t} + \rmean{v}_i\frac{\partial\rmean{\varrho\rf v\rf}}{\partial x_i} + \rmean{\varrho\rf v_i\rf}\!\cdot\!\rmean{v},_i + \rmean{v\rf v_i\rf}\!\cdot\!\rmean{\varrho},_i + \rmean{\varrho\rf v_i\rf v\rf\!,_i} + \rmean{v\rf v_i\rf\varrho\rf\!,_i} = \rmean{v}\!\cdot\!\rmean{\varrho\rf\vdrf} - \rmean{\varrho}\!\cdot\!\rmean{v\rf\vdrf}.\label{eq:edsvt}
\end{equation}
}
{
\begin{align}
\begin{split}
\frac{\partial\rmean{\varrho\rf v\rf}}{\partial t} & + \rmean{v}_i\frac{\partial\rmean{\varrho\rf v\rf}}{\partial x_i} + \rmean{\varrho\rf v_i\rf}\!\cdot\!\rmean{v},_i + \rmean{v\rf v_i\rf}\!\cdot\!\rmean{\varrho},_i\\
&\qquad + \rmean{\varrho\rf v_i\rf v\rf\!,_i} + \rmean{v\rf v_i\rf\varrho\rf\!,_i} = \rmean{v}\!\cdot\!\rmean{\varrho\rf\vdrf} - \rmean{\varrho}\!\cdot\!\rmean{v\rf\vdrf}.\label{eq:edsvt}
\end{split}
\end{align}
}
It is useful to recast these equations in the nomenclature used in moment closure of VD turbulence. Following Livescu et al. \cite{Livescu_09}, \Eqre{eq:edsvt} can be written as
\begin{equation}
\frac{\partial \Hat{b}}{\partial t} + \rmean{v}_i \Hat{b},_i = -\frac{1+\Hat{b}}{\rmean{\varrho}}(\rmean{\varrho}a_i),_i - \rmean{\varrho}\frac{\partial\rmean{v\rf v_i\rf}}{\partial x_i} + 2\rmean{\varrho}\!\cdot\!\rmean{v\rf\ivdrf},\label{eq:b}
\end{equation}
where $\rmean{v_i\ff}=-\fmean{v_i\rf}=-a_i$ and $\Hat{b}=-\rmean{\varrho\rf v\rf}$.

Moment closures, such as Ref.\ \cite{Besnard_92}, solve equations for $a_i$ and $\Hat{b}$. In \Eqre{eq:b} the last two terms represent turbulent transport and micro-mixing, respectively. Similarly, $a_i$ accounts for molecular mixing and transport, since $(\rmean{\varrho}a_i),_i = \rmean{\varrho\rf\ivdrf} + \rmean{v_i\rf\varrho\rf\!,_i}$, and thus the mass-flux term in \Eqre{eq:b} is
\ifthenelse{\boolean{JoT}}
{
\begin{equation}
-\frac{1+\Hat{b}}{\rmean{\varrho}}(\rmean{\varrho}a_i),_i = \rmean{\varrho\rf v_i\rf}\cdot\rmean{v},_i + \rmean{v\rf v_i\rf}\cdot\rmean{\varrho},_i + \rmean{\varrho\rf v_i\rf v\rf\!,_i} + \rmean{v\rf v_i\rf\varrho\rf\!,_i} + \rmean{\varrho}\frac{\partial\rmean{v\rf v_i\rf}}{\partial x_i} - \rmean{v}\cdot\rmean{\varrho\rf\vdrf} - \rmean{\varrho}\cdot\rmean{v\rf\vdrf},
\end{equation}
}
{
\begin{align}
-\frac{1+\Hat{b}}{\rmean{\varrho}}(\rmean{\varrho}a_i),_i & = \rmean{\varrho\rf v_i\rf}\!\cdot\!\rmean{v},_i + \rmean{v\rf v_i\rf}\!\cdot\!\rmean{\varrho},_i + \rmean{\varrho\rf v_i\rf v\rf\!,_i} + \rmean{v\rf v_i\rf\varrho\rf\!,_i}\nonumber\\
&\quad + \rmean{\varrho}\frac{\partial\rmean{v\rf v_i\rf}}{\partial x_i} - \rmean{v}\!\cdot\!\rmean{\varrho\rf\vdrf} - \rmean{\varrho}\!\cdot\!\rmean{v\rf\vdrf},
\end{align}
}
representing production, turbulent transport and molecular mixing.

In contrast, the left hand sides of the Eulerian equations (\ref{eq:emdtr}, \ref{eq:evdt}, \ref{eq:V} and \ref{eq:edsvt}) explicitly detail the terms originating from the physical process of advection. In the pdf framework only the terms representing small scale mixing ($\rmean{\varrho\rf\ivdrf}$, $\rmean{v\rf\ivdrf}$ and $\rmean{\varrho\rf^2\ivdrf}$ on the right hand sides) require closure assumptions. As described earlier, this is done consistently with the beta-pdf for the density, Eqs.\ (\ref{eq:im1}, \ref{eq:im2}, \ref{eq:edsv} and \ref{eq:mrvi}).

\subsection{Extension of the beta-SDE}
The extension of the SDE \Eqrs{eq:dSDEt} to inhomogeneous flows is now attempted and its derived moment equations are compared to the ones derived from mass conservation.

Compared to the homogeneous moment equations, \Eqrs{eq:ed1}, \Eqrs{eq:ed2}, \Eqrs{eq:es1}, \Eqrs{eq:ed3} and \Eqrs{eq:ed4} (with the exception of $\rmean{\varrho\rf v\rf}$) there is a single additional term in Eqs.\ (\ref{eq:ed1i}--\ref{eq:ed4i}), proportional to the mean dilatation, $\rmean{\vd}$. Interestingly, \Eqres{eq:edsv} and \Eqrs{eq:edsvi} show that the exact equation governing $\rmean{\varrho\rf v\rf}$ is the same in both homogeneous and inhomogeneous flows.

An extension of the SDE \Eqrs{eq:dSDEt} to inhomogeneous flows should satisfy the following constraints:
\begin{enumerate}
\item There must be a new drift term, otherwise the homogeneous equation governing $\rmean{\varrho}$ is not modified.
\item The new drift must be constant or linear in $\varrho^*$, otherwise the stationary solution is no longer beta.
\item The drift must have the correct unit, i.e.\ that of $\varrho^*$.
\item The drift must be tensorially correct, i.e.\ scalar.
\item There may be an additional diffusion term that is independent of, linear, or quadratic in $\varrho^*$, otherwise the solution is no longer beta.
\end{enumerate}

A possible way to account for the large scale effects in the SDE \Eqrs{eq:dSDEt}, satisfying the above constraints, is with a second drift term
\ifthenelse{\boolean{JoT}}
{
\begin{equation}
\mathrm{d}\varrho^*(t) = \left[\frac{b(\bv{x},t)}{2}(S(\bv{x},t)-\varrho^*) - \varrho^*\rmean{\vd}\right]\mathrm{d}t + \sqrt{\kappa(\bv{x},t)\varrho^*(1-\varrho^*)}\mathrm{d}W(t).\label{eq:dSDEti}
\end{equation}
}
{
\begin{align}
\mathrm{d}\varrho^*(t) & = \left[\frac{b(\bv{x},t)}{2}(S(\bv{x},t)-\varrho^*) - \varrho^*\rmean{\vd}\right]\mathrm{d}t\nonumber\\
&\quad + \sqrt{\kappa(\bv{x},t)\varrho^*(1-\varrho^*)}\mathrm{d}W(t).\label{eq:dSDEti}
\end{align}
}
The additional term represents the effect of mean volume expansion and it is straightforward to show that \Eqre{eq:dSDEti} still results in a beta-pdf, since the drift is still linear in $\varrho^*$. Note that now the three model coefficients, $b$, $S$ and $\kappa$, are assumed to depend on the spatial coordinate, $\bv{x}$, as well, allowing for spatially inhomogeneous specifications.

The Lagrangian moment equations, derived from the inhomogeneous SDE \Eqrs{eq:dSDEti}, become
{\allowdisplaybreaks
\begin{align}
\left.\ld{\rmean{\varrho}}\right|_\mathrm{sde} & = -\rmean{\varrho}\!\cdot\!\rmean{\vd} + \mathrm{hom.},\label{eq:md1i}\\
\left.\frac{1}{2}\ld{\rv{\varrho}}\right|_\mathrm{sde} & = -\rv{\varrho}\!\cdot\!\rmean{\vd} + \mathrm{hom.},\label{eq:md2i}\\
\left.\ld{\rmean{v}}\right|_\mathrm{sde} & = \rmean{v}\!\cdot\!\rmean{\vd} + \mathrm{hom.},\label{eq:ms1i}\\
\left.\ld{\rmean{\varrho\rf v\rf}}\right|_\mathrm{sde} & = \mathrm{hom.},\label{eq:mdsvi}\\
\left.\frac{1}{3}\ld{\rs{\varrho}}\right|_\mathrm{sde} & = -\rs{\varrho}\!\cdot\!\rmean{\vd} - \rv{\varrho}\!\cdot\!\rmean{\varrho}\!\cdot\!\rmean{\vd} + \mathrm{hom.},\label{eq:md3i}\\
\left.\frac{1}{4}\ld{\rk{\varrho}}\right|_\mathrm{sde} & = -\rk{\varrho}\!\cdot\!\rmean{\vd} - \rs{\varrho}\!\cdot\!\rmean{\varrho}\!\cdot\!\rmean{\vd} + \mathrm{hom.},\label{eq:md4i}
\end{align}}%
where only the new inhomogeneous terms are shown.

Comparing the first four exact equations (\ref{eq:ed1i}--\ref{eq:edsvi}) with the model equations (\ref{eq:md1i}--\ref{eq:mdsvi}), indicates that the single term, $-\varrho^*\rmean{\vd}\mathrm{d}t$, in the SDE creates the correct large scale terms, $-\rmean{\varrho}\cdot\rmean{d}$, $-2\rv{\varrho}\cdot\rmean{d}$ and $\rmean{v}\cdot\rmean{\vd}$ in the equations for $\rmean{\varrho}$, $\rv{\varrho}$ and $\rmean{v}$. However, it gives rise to a spurious term, $-n\rmean{\varrho\rf^{(n-1)}}\cdot\rmean{\varrho}\cdot\rmean{\vd}$, in the equations for $\rmean{\varrho\rf^n}$, $n\ge3$, second terms on the right hand side of Eqs.\ (\ref{eq:md3i}, \ref{eq:md4i}). Therefore, a consistent representation in \Eqre{eq:dSDEti} for inhomogeneous flows is limited to the first two moments.

This section extended the SDE \Eqrs{eq:dSDEt} for inhomogeneous flows. The representation in the SDE \Eqrs{eq:dSDEti} is consistent up to the first two density moments and for the mean specific volume and density-specific-volume covariance.

\section{Conclusion}
\label{sec:conclusion}
The rigorous mathematical consequences of assuming a beta distribution for the fluid mass density in variable-density (VD) flows have been derived. Several issues of modeling and theoretical nature have been explored in order to lay the groundwork for the related and subsequent articles \cite{Bakosi_10b,Bakosi_10c} that apply these results in model computations and validations. Our treatment have been general: treating the mixing process in terms of a general dilatational field.

The main results can be summarized as follows:
\begin{enumerate}
\item We have presented a stochastic differential equation (SDE) that yields a beta distribution and reflects conservation of mass in VD flows.
\item We have derived two sets of moment equations, one from the SDE and one from exact mass conservation, and compared them to determine to what physical processes the parameters in the SDE correspond.
\item We have drawn attention to the mean specific volume as a primary mixing quantity, necessary to close the dynamical moment equations, see Ref.\ \cite{Livescu_09}, and provided a series of relations for the correlations, $\rmean{v\rf\ivdrf}$ and $\rmean{\varrho\rf^{n}\ivdrf}$, $n\ge1$.
\item From a modeling point of view if one were to design a moment closure for the first few moments of the pdf, one can now relate the mixing statistics in the various moment equations to each other using one consistency principle: models for the mixing processes in the different moment equations are related to each other in a unique self-consistent way if the underlying pdf is beta.
\item The discussion is for arbitrary Schmidt numbers. Some results for high \textit{Sc} have been shown.
\item As an application, Appendix \ref{app:pol} shows that given the joint pdf of density (or pressure) and velocity in polytropic gases yields a closed diagnostic equation for all pressure-dilatation covariances of the form $\rmean{p\rf^n\ivdrf}$, $n\ge1$.
\end{enumerate}

Our primary interest is in developing a model that can predict material mixing in fluids with large density differences, of order 10 and larger, in flows in which the turbulence is transitional and/or non-equilibrium. We have documented several rigorous mathematical results necessary for future works. In the subsequent papers \cite{Bakosi_10b,Bakosi_10c} we apply these ideas for VD mixing in the Rayleigh-Taylor instability.

\section*{Acknowledgements}
J.\ Waltz is gratefully acknowledged for a series of informative discussions. This work was performed under the auspices of the U.S.\ Department of Energy.

\ifthenelse{\boolean{JoT}}
{\appendices}
{\appendix}

\section{Summary of results}
\label{app:sum}
The solution of the stochastic differential equation, governing the Lagrangian particle property $0\!\le\!\varrho^*(t)\!\le\!1$,
\begin{align}
\mathrm{d}\varrho^*(t) = \left[\frac{b}{2}(S-\varrho^*) - \varrho^*\rmean{d}\right]\mathrm{d}t + \sqrt{\kappa\varrho^*(1-\varrho^*)}\mathrm{d}W(t),\label{eq:AppdSDEti}
\end{align}
with $\rmean{d}=\rmean{v}_{i,i}$ and deterministic functions $b(\bv{x},t)>0$, $\kappa(\bv{x},t)>0$ and $0 < S(\bv{x},t) < 1$, is a non-stationary skewed beta distribution.

For a material mixing model in turbulent flows, the temporal evolution of the three coefficients, $b$, $S$ and $\kappa$, needs to be specified subject to the constraint
\begin{equation}
\delta = \frac{\kappa}{b} \le \frac{\rv{\varrho}}{\rmean{\varrho}(1-\rmean{\varrho})} = 1-\theta,\label{eq:Appdelta}
\end{equation}
where $\rmean{\varrho}$ and $\rv{\varrho}$ denote the mean and the variance, respectively, and $0\le\theta\le1$ is a mix-metric. The constraint \Eqrs{eq:Appdelta} ensures a non-increasing variance, as required in homogeneous flows.

In conjunction with a velocity pdf model, \Eqre{eq:AppdSDEti} for the mass density is shown to be consistent with mass conservation up to the first two density moments in a variable-density, statistically inhomogeneous, non-stationary turbulent flow.

Special cases:
\begin{enumerate}
\item Homogeneous case: The term $-\varrho^*\rmean{d}$ in \Eqre{eq:AppdSDEti} vanishes in homogeneous flows. In this case the model coefficients, $b(t)$, $S(t)$ and $\kappa(t)$, are only functions of time and \Eqre{eq:AppdSDEti} is consistent with mass conservation for all moments of the one-point physical density pdf.
\item No-mix or high-\textit{Sc} limit: Restricting $\delta\to\infty$, i.e.\ $\kappa \gg b$, results in a model that allows mixing asymmetry and allows very little molecular diffusion.
\end{enumerate}

\section{Example: Application to polytropic medium}
\label{app:pol}
Throughout the paper no restriction has been made on the relation between the state variables: the development has been independent of the equation of state. In this section we assume that the material obeys the polytropic law. We show that this, together with the assumption on the density pdf, leads to some interesting consequences. In particular, the knowledge of the joint pdf of density and velocity in polytropic gases yields a diagnostic relationship for the pressure-dilatation covariance, $\rmean{p\rf\ivdrf}$, which therefore requires no closure assumptions. $\rmean{p\rf\ivdrf}$ has been shown to be an important contributor to the budget of the turbulent kinetic energy at high turbulent Mach numbers \cite{Lele_94}.

Assuming a polytropic equation of state,
\begin{equation}
p = C\varrho^m,\label{eq:eos}
\end{equation}
with the polytropic coefficient, $m\ge0$, and a constant, $C$, from mass conservation, \Eqre{eq:mass}, one has for the instantaneous pressure
\begin{equation}
\ld{p} = -mpd.\label{eq:ep2}
\end{equation}

If the fluid density is beta-distributed, an equation for the mean pressure, $\rmean{p}(\bv{x},t)$, can be derived from the SDE \Eqrs{eq:dSDEti} and \Eqre{eq:eos}:
\ifthenelse{\boolean{JoT}}
{
\begin{equation}
\left.\frac{1}{m}\ld{\rmean{p}}\right|_\mathrm{sde} = -\rmean{p}\cdot\rmean{d} + \frac{b}{2}\left[S\left(\rmean{p}\!\cdot\!\rmean{v} + \rmean{p\rf v\rf}\right) - \rmean{p}\right] + \frac{\kappa}{2}(m-1)\left[\left(\rmean{p}\!\cdot\!\rmean{v} + \rmean{p\rf v\rf}\right) - \rmean{p}\right].\label{eq:mp1i}
\end{equation}
}
{
\begin{align}
\left.\frac{1}{m}\ld{\rmean{p}}\right|_\mathrm{sde} & = -\rmean{p}\!\cdot\!\rmean{d} + \frac{b}{2}\left[S\left(\rmean{p}\!\cdot\!\rmean{v} + \rmean{p\rf v\rf}\right) - \rmean{p}\right]\nonumber\\
&\quad + \frac{\kappa}{2}(m-1)\left[\left(\rmean{p}\!\cdot\!\rmean{v} + \rmean{p\rf v\rf}\right) - \rmean{p}\right].\label{eq:mp1i}
\end{align}
}
Comparing this to the equation derived from mass conservation, \Eqre{eq:ep2},
\begin{equation}
\frac{1}{m}\ld{\rmean{p}} = -\rmean{p}\!\cdot\!\rmean{d} - \rmean{p\rf\ivdrf},
\end{equation}
gives a diagnostic equation for the pressure-dilatation covariance as
\begin{equation}
\rmean{p\rf\ivdrf} = \frac{b}{2}\left[\rmean{p} - S\left(\rmean{p}\!\cdot\!\rmean{v} + \rmean{p\rf v\rf}\right)\right] + \frac{\kappa}{2}(m-1)\left(\rmean{p}-\rmean{p}\!\cdot\!\rmean{v} - \rmean{p\rf v\rf}\right).\label{eq:pd}
\end{equation}
All terms appearing in \Eqre{eq:pd} can be extracted from the density pdf using \Eqre{eq:eos}. For example, an equation governing $\rmean{p\rf v\rf}$ can be obtained by multiplying the FPE equivalent to the SDE \Eqrs{eq:dSDEti} by $(p-\rmean{p})(v-\rmean{v})$ and integrating each term. This yields
\begin{equation}
\left.\ld{\rmean{p\rf v\rf}}\right|_\mathrm{sde} = \mathcal{F}(b,S,\kappa,m,\rmean{p},\rmean{v},\rmean{p\rf v\rf},\rmean{p\rf v\rf^2},\dots),
\end{equation}
where $\mathcal{F}$ is a function of the SDE parameters and density and pressure statistics, which can all be extracted from the joint pdf.

The pressure variance equations, derived from the SDE \Eqrs{eq:dSDEti} and \Eqre{eq:ep2}, respectively, are
\ifthenelse{\boolean{JoT}}
{
\begin{align}
\left.\frac{1}{2m}\ld{\rv{p}}\right|_\mathrm{sde} & = -\rv{p}\!\cdot\!\rmean{\vd} + \frac{b}{2}\left(S\rmean{p\rf pv} - \rv{p}\right) + \frac{\kappa}{2}(2m-1)\left(\rmean{p\rf^2v} - \rmean{p^2}\right)\nonumber\\
&\quad + \frac{\kappa}{2}(m-1)\left(\rmean{p}^2-\rmean{p}\!\cdot\!\rmean{pv}\right),\label{eq:mp2i}\\
\frac{1}{2m}\ld{\rv{p}} & = -\rv{p}\!\cdot\!\rmean{\vd} - \rmean{p}\!\cdot\!\rmean{p\rf\vdrf} - \rmean{p\rf^2\vdrf},\label{eq:ep2i}
\end{align}
}
{
\begin{align}
\left.\frac{1}{2m}\ld{\rv{p}}\right|_\mathrm{sde} & = -\rv{p}\!\cdot\!\rmean{\vd} + \frac{b}{2}\left(S\rmean{p\rf pv} - \rv{p}\right)\nonumber\\
&\quad + \frac{\kappa}{2}(2m-1)\left(\rmean{p\rf^2v} - \rmean{p^2}\right)\nonumber\\
&\quad + \frac{\kappa}{2}(m-1)\left(\rmean{p}^2-\rmean{p}\!\cdot\!\rmean{pv}\right),\label{eq:mp2i}\\
\frac{1}{2m}\ld{\rv{p}} & = -\rv{p}\!\cdot\!\rmean{\vd} - \rmean{p}\!\cdot\!\rmean{p\rf\vdrf} - \rmean{p\rf^2\vdrf},\label{eq:ep2i}
\end{align}
}
which shows a similar trend for the pressure-dilatation covariances, $\rmean{p\rf^n\ivdrf}$, as for $\rmean{\varrho\rf^n\ivdrf}$, developed earlier: a series of relations can be derived between the SDE parameters, $b$, $S$ and $\kappa$, and the mixing physics but now expressed in terms of the pressure. The comparison of the above equations (derived from the SDE and continuity, respectively) also reveal that the extension in \Eqre{eq:dSDEti} is consistent for the first two pressure moments in inhomogeneous flows: the term $-\varrho^*\rmean{d}\mathrm{d}t$ generates the correct large scale terms, $-m\rmean{p}\cdot\rmean{d}$ and $-2m\rv{p}\cdot\rmean{d}$ in \Eqres{eq:mp1i} and \Eqrs{eq:mp2i}.

It is emphasized that the above development is a rigorous mathematical consequence of the two assumptions: 1) the density pdf is beta and 2) the medium is polytropic. The important point is that all the statistics $\rmean{\varrho\rf^n\ivdrf}$ and $\rmean{p\rf^n\ivdrf}$, $n\ge1$ are known in a polytropic medium given the joint pdf of density and velocity.

To summarize:
\begin{enumerate}
\item In polytropic media the joint pdf of density (or pressure) and velocity provides a series of consistent relations for the correlations, $\rmean{\varrho\rf^n\ivdrf}$ and $\rmean{p\rf^n\ivdrf}$ for $n\ge1$, in terms of the parameters of the pdf.
\item Assuming a beta-pdf for the density, the equations governing the first two pressure moments in polytropic gases have been derived.
\item The inhomogeneous extension of the density-SDE \Eqrs{eq:dSDEti} consistently represents the first two moments of the pressure.
\end{enumerate}

\section{Symmetric case: $S=1/2$}
\label{app:sym}
A special case of the beta-pdf can be obtained if only a symmetric distribution is allowed. Physically, this corresponds to mixing of two equal amounts of scalars with equal diffusivity.

This case is obtained by setting $S=1/2$, which results in the following stationary values for the first four moments, deduced from \Eqres{eq:sdmean}, \Eqrs{eq:sdv}, \Eqrs{eq:ss} and \Eqrs{eq:sk},
{\allowdisplaybreaks
\begin{align}
\rmean{\varrho}^\mathrm{sym}_\mathrm{s} & = \frac{1}{2},\\
\rv{\varrho}^\mathrm{sym}_\mathrm{s} & = \frac{1}{4}\frac{\delta}{1+\delta},\\
\left.\frac{\rs{\varrho}}{\rv{\varrho}^{3/2}}\right|^\mathrm{sym}_\mathrm{s} & = 0,\\
\left.\frac{\rk{\varrho}}{\rv{\varrho}^2}\right|^\mathrm{sym}_\mathrm{s} & = \frac{1+\delta}{1/3+\delta}.
\end{align}}%
This set of equations is characteristic of the shape of the pdf in the symmetric case as mixing progresses. At any time the mean converges to $1/2$, the variance decays with $\delta(t)$, the skewness decays to zero and the kurtosis increases from $1$ to $3$.

The model governing equations for the first four moments can be obtained by setting $S=1/2$ in \Eqres{eq:dmean}, \Eqrs{eq:dv}, \Eqrs{eq:3m} and \Eqrs{eq:4m}, resulting in
\begin{align}
\frac{1}{b}\left.\frac{\partial\rmean{\varrho}}{\partial t}\right|^\mathrm{sym} & = \frac{1}{2}\left(\frac{1}{2}-\rmean{\varrho}\right),\\
\frac{1}{b}\left.\frac{\partial\rv{\varrho}}{\partial t}\right|^\mathrm{sym} & = \frac{\delta}{4} - \left(1+\delta\right)\rv{\varrho},\\
\frac{1}{3b}\left.\frac{\partial\rs{\varrho}}{\partial t}\right|^\mathrm{sym} & = -\left(\frac{1}{2}+\delta\right)\rs{\varrho},\\
\frac{1}{2b}\left.\frac{\partial\rk{\varrho}}{\partial t}\right|^\mathrm{sym} & = \frac{3}{4}\delta\rv{\varrho} - \left(1+3\delta\right)\rk{\varrho}.
\end{align}
These equations show the effect of molecular diffusion in the special case of symmetric mixing. The time evolution of the single model parameter, $\delta(t)$, fully determines all moments of the pdf. The model coefficient $b(t)$ acts as the mixing rate: the full right hand sides are multiplied by multiples of $b$.

Setting $S=1/2$ in Eqs.\ (\ref{eq:mi1}--\ref{eq:mi3}) gives the closures for the density-velocity-dilatation covariances, $\rmean{\varrho\rf^n\ivdrf}$, in the symmetric case as
\begin{align}
\rmean{\varrho\rf\vdrf}^\mathrm{sym} & = 0,\label{eq:mi1-sym}\\
\rmean{\varrho\rf^2\vdrf}^\mathrm{sym} & = \frac{b}{2}\left[(1+\delta)\rv{\varrho} - \frac{\delta}{4}\right],\label{eq:mi2-sym}\\
\rmean{\varrho\rf^3\vdrf}^\mathrm{sym} & = \frac{b}{2}\left[\frac{\delta}{8} - \frac{1}{2}(1+\delta)\rv{\varrho} + (1+2\delta)\rs{\varrho}\right].\label{eq:mi3-sym}
\end{align}
Eqs. (\ref{eq:mi1-sym}--\ref{eq:mi3-sym}) show the effect of molecular mixing on the statistics $\rmean{\varrho\rf^{n}\ivdrf}$, governed by the single model parameter, $\delta(t)$, in the symmetric case. Representing molecular diffusion requires the knowledge of all the central moments, $\rmean{\varrho\rf^{n}}$, to obtain $\rmean{\varrho\rf^{n}\ivdrf}$, $n\ge2$.

\ifthenelse{\boolean{JoT}}
{\bibliographystyle{tJOT}}
{\bibliographystyle{physfluids.bst}}
\bibliography{jbakosi}

\end{document}